\documentclass[11pt,a4paper]{article}
\usepackage{jcappub}
\usepackage[utf8]{inputenc}
\usepackage{amsmath, amsthm, mathtools}
\usepackage{hyperref}
\usepackage{xcolor}
\usepackage[normalem]{ulem} % use normalem to protect \emph
\newcommand\hl{\bgroup\markoverwith
  {\textcolor{yellow}{\rule[-.5ex]{2pt}{2.5ex}}}\ULon}

\newcommand\rhl{\bgroup\markoverwithHahn2013
  {\textcolor{orange}{\rule[-.5ex]{2pt}{2.5ex}}}\ULon}
  
 \newcommand\ghl{\bgroup\markoverwith
  {\textcolor{green}{\rule[-.5ex]{2pt}{2.5ex}}}\ULon} 
\title{Identifying Dark Matter Haloes by the Caustic Boundary}

\author[a]{Sergei F. Shandarin}
\affiliation[a]{Department of Physics and Astronomy, University of Kansas\\
1251 Wescoe Hall Drive \#1082, Lawrence, Kansas 66045, U.S.A.}
\emailAdd{sergei@ku.edu}

\abstract{
Dark matter density is formally infinite at the location of caustic surfaces, where dark matter sheet folds in phase space. 
The caustics separate multi-stream regions with different number  of streams. Volume elements change the parity  
by turning inside out when passing through the caustic stage. 
Being measure-zero structures, identification of caustics via matter density fields is usually restricted 
to fine-grained simulations.
Instead a generic purely geometric algorithm can be employed to identify caustics directly by using  triangulation 
of  Lagrangian sub-manifold ${\bf x}({\bf q},t)$ where ${\bf x}$ and ${\bf q}$ are Eulerian and Lagrangian coordinates
obtained in N-body simulations.
The caustic surfaces are approximated by a set of triangles with vertices being the particles  in the simulation.  
It is demonstrated  that finding a dark matter halo is quite feasible by building its outermost convex caustic.
Neither more specific assumptions  about the geometry of the boundary nor ad hoc parameters are needed.  
The halo boundary  in our idealized but undoubtably generic simulation is neither spherical nor ellipsoidal
but rather remarkably asymmetrical. The analysis of the kinetic and potential energies of individual
particles and the halo as a whole  along with an examination of the two-dimensional phase space has
shown that the halo is gravitationally bound. }


\begin{document}
\maketitle

%-------------------------------------------------------------------------------------------------------------------------------

\section{Introduction}  %                <=============================================               sec1
\label{sec:intro}
Caustics along with the multi-stream regions and flip-flop field are  inherent features of the cold collision-less 
dark matter (DM) web.
Known in geometric optics as an  envelope of light rays reflected or refracted by a smooth curved surface
a caustic is a line or point where the light intensity approaches infinity when the wavelength of the illuminating light tends to zero.
The Zeldovich Approximation (ZA) \citep{Zeldovich1970} in two-dimensional space is identical to refraction of 
parallel light rays by a  plate with thickness given by a two-dimensional  random smooth function. 
When  a screen is placed behind the plate then bright  caustic patterns can be observed 
in a certain range of distances from the plate. These patterns are exactly the same as 
the web structures predicted by the ZA in two-dimensional case  \citep{Zeldovich1983,Shandarin1989}.  
The intensity of light in caustic is high but  finite since the wave nature of light. 
 Similarly the density in caustics in a cold collision-less DM is high but finite because of two reasons.
First,  DM  is not a continuous medium and second -- more importantly -- because a finite thermal velocity dispersion.
However, the smaller the velocity dispersion the higher  density in the caustics, see e.g. \citep{Zeldovich_Shandarin:82}.     %%%%%%%%
Nevertheless the approximation of DM by a cold continuous medium is extremely  accurate and commonly used
in cosmology.

Along with the caustics the CDM web possesses two additional  traits that cannot be found in a collisional medium like baryonic 
component in the universe. They are multi-stream flows and  flip-flops of fluid particles. All three closely connected but 
not identical phenomena. They can be used as additional quantitative characteristics of the DM web.
The multi-stream field is simply a count of the  streams with distinct velocities at every point in Eulerian space.
Generally the number of streams is an odd integer  except the caustic surfaces where it is an even integer.
The flip-flop field is the count of turns  inside out for each fluid element. Both can be evaluated in cosmological N-body 
simulations.  It can be done either on particles or tetrahedra of the tessellation of the three-dimensional  phase space 
sheet  in six dimensional  phase space \citep{Shandarin2012, Abel2012}. 
The tessellation technique allows to considerably improve N-body simulations \citep{Hahn2013,Hahn-etal:15,Hahn2016a}
and provides additional effective  diagnostics for the analysis of the complexity of the DM web  \citep{Shandarin2014,Ramachandra2015,Ramachandra17a,Ramachandra17b,Shandarin2016}.  

The multi-stream field in N-body simulations  is naturally to evaluate 
on arbitrary set of diagnostic points  in Eulerian space  \citep{Shandarin2011,Shandarin2012,2018MNRAS.477.3230S}. 
On the other hand  the flip-flop  field can be  easily computed on particles or  tetrahedra in Lagrangian space
  \citep{Shandarin2014,Shandarin2016}. Of course
the flip-flop field can be  mapped to Eulerian space but this would significantly change its mathematical properties
-- being a field in Lagrangian space -- it becomes  a multi-valued function in Eulerian space which is more difficult
to deal with. 
Caustics are the interfaces between regions with different values of flip-flops in Lagrangian space and between regions 
with different values of the number of streams in Eulerian space.

 In principle  caustics surfaces can be identified as very thin layers of very high DM densities but it would require
N-body simulations with unfeasibly high mass resolution  for realistic cosmological simulations.
However, since caustics separate the Lagrangian neighbor elements with different number of flip-flops
 it is also possible to use the common faces of two neighboring tetrahedra with  opposite parities
as a natural representation of the elements of caustic surfaces \citep{Shandarin2012,Abel2012}.

%This paper is focused on the geometrical and topological properties of DM caustics. 
Topological analysis and classification of all  generic types of caustics originating in a potential mapping of
a collision-less medium in two and three dimensions were provided in \cite{Arnold1982b}.
However, the analysis was based on the so called normal forms which roughly speaking 
are the minimal polynomials used as generators of singularities. In this form the results can be used only
as a solid guideline for the much more strenuous analysis of realistic DM flows in cosmological simulations.
The first  analysis of the geometry and topology of the caustic structures in the frame of the ZA with 
smooth random initial perturbations was done in \cite{Arnold1982}, however, it was limited to two dimensions.

For the recent  scrutiny of  the subject see \cite{Hidding2014} and \cite{Feldbrugge2018}.
However, the authors of \cite{Hidding2014} showed the 3D caustics only in  the ZA simulation
and the authors of \cite{Feldbrugge2018} showed  bicaustics (according to Arnold's terminology) in 3D N-body simulations
but not the caustics themselves. In simple terms, a bicaustic is the trace of an instantaneous caustic moving with time. 
In this paper we focus on A2 caustics which are surfaces at each instant of time after 
emergence at the nonlinear stage.

The presence of caustics has been implied in most if not all modern cosmological N-body simulations.
For example, a {\it splashback} radius defined in \cite{More2015, Mansfield2017} 
as the distance from the center of a halo to its outermost closed caustic in the spherical models.
% as an alternative to virial radius for defining boundaries of DM haloes.
The authors argued that the splashback  radius is a more physical  choice of the halo's boundary than one
based on a density contrast $\Delta$ relative to a reference  density (mean or critical).
However, they did not identify caustics directly.  Instead they searched for a minimum
of the logarithmic slope of the spherically averaged density profiles of the haloes. 
Spherical averaging of  the density profiles may
enhance the robustness of the results but it imposes an  assumption that haloes can be well described 
as spherical configurations. 
%Direct identification of the outermost caustic in simulation makes this assumption unnecessary.  
We show how to find the outermost closed caustic without making assumption of spherical symmetry.

For the first time we analyze the internal structure of a filament and halo defined by a caustic boundary. 
In particular, we focus  on ubiquitous structures forming a sequence of haloes linked by filaments.
They are clearly seen in the density distributions  shown in \citep{Kaehler-etal12, More2015}.
 The density field was rendered from dark matter  simulations using the tetrahedral  tessellation approach. 
 In both cases the images were generated with the new rendering method based on the full tetrahedral cell-projection  
 developed in  \cite{Kaehler-etal12}. 
 We have examined a basic element of this structure consisting of two haloes defined as the interiors of two closed 
 caustic surfaces linked by a quasi-cylindrical caustic  in our  CDM simulation. 
 The structure is resembling a dumbbell, therefore it will be referred to as a dumbbell structure. 
 Attaching a filament to one of the haloes then a halo to the end of the attached filament
 makes a longer structure consisting of three haloes and two filaments. These type of assembly is
 also present in abundance in figure 1 in \cite{Kaehler-etal12}.

It has been long known  that sampling along with the mass and force resolutions plays a crucial role in delineating  the internal geometry 
of the filaments and walls.
%It is determined by the number of points used  to represent the corresponding structure.
The importance of having a sufficiently high density of mass tracers for disclosing the structures was demonstrated  in 2D N-body simulations  \cite{Melott1989,Melott1990}.  In particular,  they compared the structures obtained in the simulations with physically identical initial conditions
but with different number of mass tracers. A plot with  $256^2$ particles demonstrated complicated internal structures
 arising at late nonlinear stages. 
However,  these structures were practically invisible in the  plots with  $64^2$ particles.

The studies of the properties of cosmic velocity fields in a collision-less medium especially the multivalued character of 
the flows easily reveal the discontinuities at caustics, see e.g. 
 \cite{Shandarin2011,Vogelsberger2011,Shandarin2012,Abel2012,Hahn2013,Shandarin2014,Hahn-etal:15,Shandarin2016,2018MNRAS.477.3230S}.
 However, we are not familiar with attempts of explicit construction of A2 caustic surfaces in N-body simulations
 in three-dimensional space.  Therefore  we examine the velocity fields in separate streams within a halo and filament. 
 
The studies of caustics are often conducted in the context of indirect detecting of dark matter,  
e.g. \cite{2006MNRAS.366.1217M, 2018MNRAS.477.3230S}.  However, the caustics are the surfaces
where the velocity fields experience extraordinary metamorphoses: fluid elements go through each other
and turn inside out. Caustics provide natural boundaries between different states of a collision-less medium

The major purpose of this paper  is to analyze this type of structure identified  by means of physical attributes only: 
by multi-streams, flip-flop fields and caustic surfaces without invoking any ad hoc assumptions or  parameters.
The major difficulty of this program is the requirement of large density of mass tracers. 
The simplest way to overcome this problem is to apply the approach of \cite{Melott1989,Melott1990}  
to three-dimensional  case.

%We will discuss the major difficulties and suggest some approaches to their solution. 

In section \ref{sec:ZA}, we briefly discuss caustic formation in the context of the Zeldovich approximation. 
We explain our choice of the parameters in our N-body simulation in section \ref{sec:NB}.
Section \ref{sec:Method} describes the details of our algorithm using the Lagrangian tessellation scheme. 
In section \ref{sec:CaustHigh}, we carry out the analyses of the  caustic surfaces and their relation to multi-stream field.
Section \ref{sec:dumbbell} describes the dumbbell structure and discuss the velocity field within it.
Finally we summarize the results in section \ref{sec:summary}.

\section{Zeldovich Approximation and singularities}  %          <==================================           sec2
\label{sec:ZA}
In this section we introduce the concept of singularities in a {\it cold continuous  collision-less} medium.
All three requirements are necessary and sufficient for the formation of  singularities. Cold dark matter (CDM) is an almost perfect
example of such a medium. We begin with a brief illustration by describing the evolution of CDM density field
using the Zeldovich approximation.
The ZA is an elegant analytical technique to describe the early phase of the non-linear stage of the growth of density perturbations.
 Technically it is a first order Lagrangian perturbation theory known as LPT1. However, Zeldovich suggested to extrapolate it to the beginning of the non-perturbative nonlinear stage and predicted the formation of caustics which are the boundaries of the first very thin multistream regions dubbed by him as `pancakes'. The ZA describes a dynamical mapping from the initial Lagrangian space with coordinates $\mathbf{q}$ to Eulerian space $\mathbf{x}(t)$ at time $t$. In comoving coordinates, $\mathbf{x} = \mathbf{r}/a(t)$ where $a(t)$ is the scale factor normalized by $a(z=0)=1$ and $\mathbf{r}$ are the physical coordinates of particles at time $t$ the ZA takes the form
\begin{equation} \label{eq:ZA1}                   %       <=========================================                Eq. 1
 \mathbf{x}(\mathbf{q}, t ) = \mathbf{q} + D(t)\, \mathbf{s}(\mathbf{q}),
\end{equation}
where $D(t)$ is the linear density growth factor. The potential vector field $\mathbf{s(q)} = - \nabla_q \psi(\mathbf{q})$ is determined 
by the potential $\psi(\mathbf{q})$ which is proportional to the gravitational potential at the linear stage. 
Conservation of mass implies $\rho(\mathbf{x}, t) \,d^3x = \rho_0 \,\,d^3q $, so the density field  in terms of Lagrangian coordinates is given at $t>0$ as 
\begin{equation} \label{eq:J}  %                <==============================================                     Eq 2
 \rho(\mathbf{q}, t) = \rho_0 \left| J \left[ \frac{\partial\mathbf{x}}{\partial\mathbf{q}} \right] \right|^{-1},
\end{equation}
where the Jacobian $J \left[ \frac{\partial\mathbf{x}}{\partial\mathbf{q}} \right]$ is calculated by differentiation of  Equation \ref{eq:ZA1}. Moreover, diagonalization of the symmetric deformation tensor $d_{ij} = - \nabla_q \mathbf{s(q)} =  \partial^2 \psi(\mathbf{q})/ \partial q_i \partial q_j$ in terms of its eigenvalues $\lambda_1(\mathbf{q})$, $\lambda_2(\mathbf{q})$, $\lambda_3(\mathbf{q})$ particularizes the patterns of collapsing 
of the fluid elements. This reduces the equation describing the  mass density to a convenient form
\begin{equation} \label{eq:density} %            <=========================================                        Eq 3
 \rho(\mathbf{q}, t) = \left|  \frac{\rho_0}{ \left[1 - D(t) \lambda_1(\mathbf{q}) \right]\left[1 - D(t) \lambda_2(\mathbf{q}) \right]\left[1 - D(t) \lambda_3(\mathbf{q}) \right] } \right|.
\end{equation}

Since the deformation tensor $d_{ij}$ and its eigenvalues depend only on the initial  fields, the ordered eigenvalues defined in Lagrangian space $\lambda_1(\mathbf{q}) > \lambda_2(\mathbf{q}) > \lambda_3(\mathbf{q})$ determine collapse condition for fluid elements in Eulerian space. 
With the growth of $D$ with time, the mass density of cold continuous 
fluid can rise until reaching singularity at $D(t) = 1/\lambda_1(\mathbf{q})$.  

Locally in Lagrangian space, the condition $\lambda_1(\mathbf{q})  = 1/D(t)$ firstly  takes place
at a maximum of $\lambda_1(\mathbf{q}) = max = 1/D(t_{b}) $  where $t_{b}$ denotes the time of the `pancake's birth'.
At later times  $t > t_b$ the caustics are the level surfaces %where $\lambda_1(\mathbf{q})$ takes a given 
of constant value of $\lambda_1(\mathbf{q}) =1/D(t) < 1/D(t_b)$ mapped to Eulerian space by Equation \ref{eq:ZA1}.  
At small $\delta t = t - t_b$ the level  surfaces are closed and convex in Lagrangian space.
In Eulerian space at time $t$ they form the surfaces where  density becomes formally infinite
%if the continuous  medium approximation  is used
 - therefore the term a caustic.

Mapping of the interior region within a such surface into Eulerian space  by Equation \ref{eq:ZA1}  turns it inside out 
which is possible only  in a collision-less medium.
The first collapse compresses such a region into  a very thin layer -- Zeldovich's pancake --
which consists of three  overlapping streams moving with  distinct velocities.
The first eigenvalue is greater than the boundary value $\lambda_1(\mathbf{q}) >1/D(t)$ in the interior
region only and therefore the  Jacobian $J(\mathbf{q},t)$ in Equation \ref{eq:J} is negative in the corresponding stream. 
This happens because the mapping from Lagrangian to Eulerian space results in forming folds in  phase space 
$\mathbf{x}(\mathbf{v},t)$  as well as in  $\mathbf{q}(\mathbf{x},t)$ space. It is worth mentioning that
$\mathbf{x}(\mathbf{q},t)$ is a field i.e. a single valued vector function at any time.
In contrast the  both  $\mathbf{v}(\mathbf{x},t)$ and its inverse $\mathbf{x}(\mathbf{v},t)$ remain multivalued in multi-stream regions.
Thus the first pancakes are always the regions of three-stream flows  bounded 
by a closed caustic surface a part of which is convex while the other part is concave. 
The interface of the convex and concave parts of A2 surface is a cuspidal wedge. 
The caustic surface separates the regions with  $J(\mathbf{q},t)$   having opposite signs
 in Lagrangian space while in Eulerian space it separates
the pancake from the  single-stream field surrounding the pancake. 

At a randomly chosen point  $\mathbf{q}$ the three eigenvalues are always  distinct from each other
if the perturbation $\psi(\mathbf{q})$  is a generic field.
Three fields   $\lambda_1(\mathbf{q}),\lambda_2(\mathbf{q})$ and $\lambda_3(\mathbf{q})$  associated 
with three eigenvalues
are non-Gaussian even when the generating potential  $\psi(\mathbf{q})$ is a Gaussian field. 
In the case of the Gaussian potential the joint PDF of three eigenvalues can be found in analytic form 
(see e.g. \cite{Doroshkevich1970} and \cite{Lee1998}). It contains a factor
 $(\lambda_1 - \lambda_2) (\lambda_1 - \lambda_3) (\lambda_2 - \lambda_3)$ which explicitly
  shows that the chance of finding a point with  two equal eigenvalues equals  zero.
  %}
  
However, it is worth stressing that  the points with two equal eigenvalues  exist but the  points with all
three equal eigenvalues do not. The points of equality of only two eigenvalues ($\lambda_1 = \lambda_2$ or 
$\lambda_2 = \lambda_3$) make up lines that are sets of measure zero in three-dimensional space 
thus the zero probability of finding them by a random search. The lines occur because each of the above equations
actually  hides two equations. At the point of equality of two eigenvalues (say $\lambda_1 = \lambda_2$) 
two corresponding eigenvectors are degenerate in a plane that is orthogonal to the third eigenvector.  
This means that the corresponding minor in the deformation tensor is diagonal in any coordinate system in this plane.
This can be satisfied only if $d_{11} = d_{22}$ and $d_{12}=d_{21} = 0$.
It is worth stressing that despite of the equality of two eigenvalues the collapse is  non cylindrical even locally.
These lines can be considered as the progenitors of the first filaments.

A claim  $\lambda_1(\mathbf{q}) = \lambda_2(\mathbf{q}) = \lambda_3(\mathbf{q})$  actually requires to satisfy 
five conditions simultaneously. 
If such a point $\mathbf{q}(\lambda_1=\lambda_2=\lambda_3 = \Lambda)$ existed than 
a symmetric deformation tensor $d_{ik}= \partial^2{\psi} / \partial{q_i}\partial{q_k}$  must be diagonal 
 in arbitrary Cartesian system with all diagonal elements $d_{11}=d_{22}=d_{33} =\Lambda$ 
 and three off-diagonal elements must be equal to zero $d_{12}=d_{13}=d_{23}=0$.
As a result the set of five equations  for three unknown coordinates $(q_1, q_2, q_3)$  is overdetermined 
and thus has no solution.

For more detailed analysis of the geometry and topology of the caustic structures in  two-dimensional case we
refer to \cite{Arnold1982} and \cite{Hidding2014}.  
Unfortunately a detailed analytical characterization of 3-dimensional ZA with generic initial perturbations has not reach 
a similar level yet,  however, an important step forward was recently made   in  \cite{Feldbrugge2018}.

The geometrical and topological complexities of the caustic surfaces  in three-dimensional configuration space 
 are due to the maze-like map of the three-dimensional hypersurface $\mathbf{x}(\mathbf{q})$ called the Lagrangian submanifold from six-dimensional $(\mathbf{q},\mathbf{x})$-space to three dimensional $\mathbf{x}$ space \citep{Shandarin2012,Abel2012}. The Lagrangian submanifold is a single valued, smooth and differentiable vector function 
 $\mathbf{x}=\mathbf{x}(\mathbf{q})$.
 % However, its inverse i.e. $\mathbf{q}=\mathbf{q}(\mathbf{x})$ is generally 
 %multivalued vector function similar to $\mathbf{v}=\mathbf{v}(\mathbf{x})$.
 The projection on to 3-dimensional Eulerian space is entangled with creases, kinks and folds. Note that this  submanifold is very different from one in phase space 
$(\mathbf{x},\mathbf{v})$,  though they are connected by a canonical transformation. Delineating the Lagrangian submanifold reveals several properties of the dark matter dynamics not inferred from position-space analyses. 
Two physically  related fields -- the multistream field $n_{\rm str}(\mathbf{x})$ in Eulerian space \citep{Shandarin2011,Shandarin2012, Ramachandra2015,Ramachandra17a,Ramachandra17b} 
and the flip-flop field $n_{\rm ff}(\mathbf{q})$ in Lagrangian space  %Shandarin \& Medvedev 2017, 
\citep{Shandarin2016} % n_{\rm str}
are instrumental for the analysis of the Lagrangian submanifold. 
We have already mentioned that both $n_{\rm str}(\mathbf{q})=n_{\rm str}(\mathbf{x}(\mathbf{q}))$ and 
$n_{\rm ff}(\mathbf{x})=n_{\rm ff}(\mathbf{q}(\mathbf{x}))$ are multi valued functions.

Equation \ref{eq:density} shows that the maximum number of flip-flops experienced by a fluid element can be only
in the range from zero to three.
These are determined by four conditions imposed on the eigenvalues:
 $\lambda_1 <0$ or  $\lambda_1 >0$ or  $\lambda_2>0$  or  $\lambda_3 >0$ respectively. 
 This represents a fundamental limitation of the ZA. 
The number of flip-flops explicitly evaluated  in cosmological N-body simulations 
currently does not exceed  a few thousand \citep{Vogelsberger2011, Shandarin2016}. 
The directly registered  number of streams 
does not exceed $\sim10^{5}$   \cite{Abel2012}. However,   indirect estimates  
predict  $\sim10^{14}$  streams at a typical point at 8 kpc from the halo center of DM Milky Way haloes
simulated in  the Aquarius Project \cite{Vogelsberger2011}. 
The number of flip-flops is always less than the number of streams. For instance, consider the one-dimensional collapse of
a sinusoidal density perturbations. 
The fluid elements of the central part of the halo experience the greatest number of flip-flops at all times.
If the maximum of flip-flops is $n_{\rm ff,max}$  then the maximum of the number of streams 
becomes $n_{\rm str,max} = 2n_{\rm ff,max}+1$.  
In addition the process of merging of haloes may considerably increase the number of streams without 
significant growth of flip-flops  in individual streams. Unfortunately, there is no simple relation between numbers of flip-flops and streams in a general case.

\section{N-body simulation} %        <====================================           sec3
\label{sec:NB}
We  carried out  our simulations  with the standard $\Lambda$CDM cosmology, 
$\Omega_m=0.3,~\Omega_\Lambda=0.7,~ \Omega_b=0, ~\sigma_8=0.9,~ h=0.7$.
 In order to compute flip-flops we used a slightly modified version (for details see \cite{Shandarin2016}) 
of a publicly available cosmological TreePM/SPH  code GADGET \citep{Springel:05}. 

In order to have sufficient particle sampling  we introduce an artificial  spherically symmetric sharp cutoff 
in the initial spectrum of perturbations -- a standard option in GADGET. 
Our simulations differed from each other by four  parameters: number of particles $N^3$,   the size of the box $L$ and
the force  softening scale $R_{\rm s}$  both in units of $ h^{-1}$ Mpc, 
and a cutoff scale $k_{\rm c}$  defined in GADGET by the parameter  $n_{\rm c}$  in the  equation   
$k_{\rm c} =  ( 2 \pi /L) (N / n_{c})$. 
In other words it means that the initial power spectrum $P(k) = 0$ for all $k >k_{\rm c}$.

After trying a number of different sets of the parameters  we have selected one that fits our major goals.
The size of the box is $L = 100 h^{-1}$ Mpc, the number of particles $N=256^3$, the force softening scale $R_{\rm s}=0.8h^{-1}$ ~Mpc,
and the cutoff scale parameter $n_{\rm c}=64$ which means that the initial spectrum covers a very small range from $k_{\rm min}= ( 2 \pi /L)$ to 
$k_{\rm max}= k_{\rm c}=4( 2 \pi /L)$. 
The choice of $R_{\rm s}$  about two times greater than a mean particle separation $\overline{d} \approx L/N = 0.39 h^{-1}$ Mpc 
has been made in order to preserve the  phase space sheet from self-crossing for longer period of time.
As it was demonstrated  in the one-dimensional oblique plane wave collapse  test the discrepancy between 
the 3D N-body solution and the true one-dimensional solution already exists at shell-crossing and becomes 
more severe at later times already at $R_{\rm s} = 0.5  \overline{d}$  \citep{Melott1997}.
 Later this result was confirmed and  elaborated by additional more sophisticated  tests in \cite{Angulo2013a} and \cite{Hahn2016a}.
Increasing   $R_{\rm s}$  in our simulations  to $R_{\rm s} \approx 2 \overline{d}$ significantly relieves the problem 
of the self-crossing of the phase space sheet. 
The initial power spectrum of this idealized simulation is similar to that in HDM simulations in $100 h^{-1}$ Mpc 
box by \cite{2007MNRAS.380...93W}.  
The major difference is at $k \gtrsim 0.1 h$ Mpc$^{-1}$: in our simulation $P(k) =0$ while
in the HDM model $P(k)$ falls exponentially so that $P(k)$  drops off almost seven times at $k = 0.2 h$ Mpc$^{-1}$. 
The other difference is in the force softening length: it was about twenty times less than in our simulation.

The cutoff in the initial power spectrum requires slightly more than a hundred random numbers for the generation
of the initial perturbations. The number is obviously far too small for any statistical valuation but it is more than sufficient 
to guarantee  the initial conditions to be of a generic type. 

Our idealized model allows a reliable identification of several generations of internal caustics.  We also were able 
to simulate one of the most fundamental structures in DM web. It is a dumbbell structure consisting of two haloes connected by a cylindrical filament. As we mentioned in section \ref{sec:intro}  numerous images of this structure are shown 
in  figure 1 in \cite{Kaehler-etal12}.
However, due to a small range of the initial power spectrum this simulation can be regarded only as the first successful attempt
to directly build the caustic surfaces in cosmological N-body simulation started from a smooth random field which guaranties
{\it only that  the  identified caustics are of generic types}. 
Our choice of the cutoff scale roughly correspond to the scale of massive clusters of galaxies. 
Very roughly it might qualitatively illustrate the formation of clusters of galaxies in the HDM scenario and perhaps
first haloes in WDM and CDM models.
\begin{figure} 	%	     <======================================================                   fig1	
\centering\includegraphics[width=6.5cm]{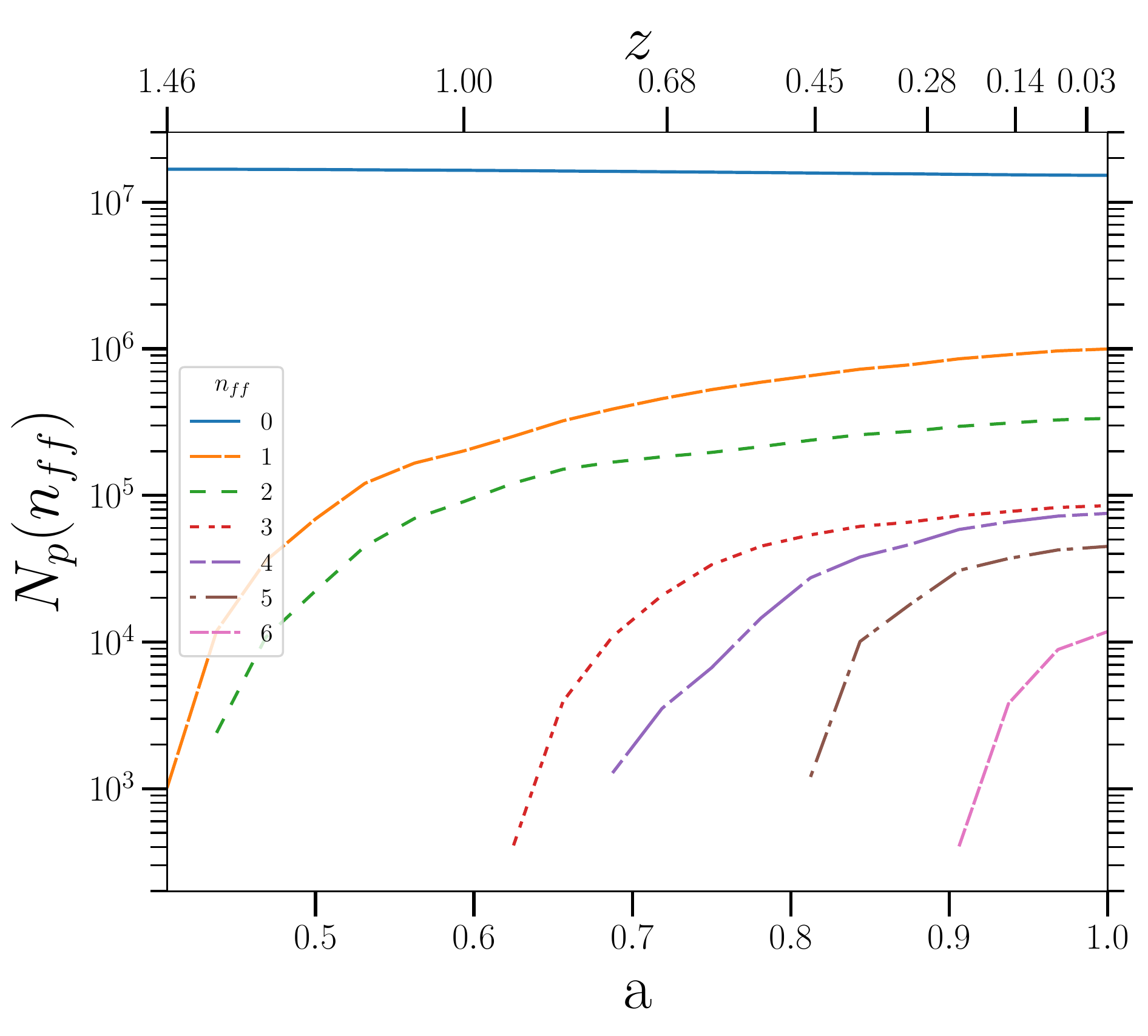}
\caption{Number of particle $N_{\rm p}$ that experienced $n_{\rm ff}$ flip-flops as a function of scale factor $a$.
Redshift is shown on the top axis. }
\label{fig:ff-evolution}
\end{figure}

%Figure \ref{fig:ff-evolution} gives some idea on the number of particles involved in building the multi-stream structure
%and caustics.
%It shows the growth of the number of particles in the structure as a function of scale factor.

Figure \ref{fig:ff-evolution} provides a sense of the evolution of the structure in the simulation at the nonlinear stage.
 It shows the monotonic growth of the number of particles experienced flip-flops and the decrease of the number of particles
 with zero flip-flops $N_p(n_{\rm ff},a)$. 
 It is also worth noting  an orderly behavior of $N_p(n_{\rm ff},a)$:  $N_p(n_{\rm ff},a) > N_p(n_{\rm ff}+1,a)$ for all $a$ and 
 $N_p(n_{\rm ff},a_2) > N_p(n_{\rm ff},a_1)$  for all $n_{\rm ff}$  if $a_2 > a_1$.
  The number of particles drops from $\approx 10^6$ at $n_{\rm ff} =1$ to $\approx 10^4$ 
 at $n_{\rm ff} =6$. It is worth stressing that the number of vertices in the caustics in the region we discuss in section \ref{sec:CaustHigh}
 is 10--100 times less as Table 1 in Appendix A indicates. 

%For instance  two red compact shells shown in Figures \ref{fig:full-E}, \ref{fig:red-bulbes}, \ref{fig:center-E}, and \ref{fig:full-E}  
%-- which we consider as two examples of splashback caustics --  are composed of about 30000 triangles and 24000 vertices 
%combined.

\section{Identification of caustics in numerical simulations} %                   <========================           sec4
\label{sec:Method}
Finding caustics in numerical simulations by the ZA as well as by N-body  techniques could be 
made by various methods.  One can do this by analyzing singularities of  the eigenvalue fields 
of the deformation tensor $\partial s_i /  \partial q_j$,
where $s_i(\mathbf{q},t) = x_i(\mathbf{q},t) - q_i$. In the case of the ZA it is simply $D(t)s_i(\mathbf{q})$ 
in Equation \ref{eq:ZA1}.
In the case  of N-body simulations  it requires numerical calculation of the positions of the particles at time $t$.
The passage of a particle through the singular stage can be  registered through the change of sign
of the Jacobian $J$ (Equation \ref{eq:J}), see e.g.  \cite{Vogelsberger2011, Shandarin2016}.  
This method  allows to count how many of times a particle passed through the caustic state which is equivalent
to the number of  flip-flops computed for each particles.
 The probe of the geometry and topology of caustics
requires to analyze the spacial structure of the eigenvalue fields  \cite{Arnold1982,Hidding2014, Feldbrugge2018}.

In this paper we will use a different method of finding caustic surfaces directly from the Lagrangian triangulation \citep{Shandarin2012,Abel2012}.
First we need to evaluate the volume of every tetrahedron in the tessellation.
It can be done  by computing the following determinant for each tetrahedron
\begin{equation} 
\label{eq:det}
d = 
\begin{vmatrix}
x_{1} & y_{1} & z_{1} & 1 \\ 
x_{2} & y_{2} & z_{2} & 1 \\ 
x_{3} & y_{3} & z_{3} & 1 \\ 
x_{4} & y_{4} & z_{4} & 1 
\end{vmatrix},
\end{equation}
where  $x_i, y_i, z_i$ are the coordinates of four vertices of the tetrahedron. It is easy to see that the determinant $d$ can be
either positive or negative because changing the order of the vertices results in swapping a pair of rows in the determinant
resulting in the change of its sign.
Each determinant $d$ in the tessellation can be made positive at the initial time.  
Thus the volume of every tetrahedron becomes $V_i = d_i/6$.
Similarly to the Delaunay triangulation, the Lagrangian algorithm considers the particles as  the tracers of collision-less DM fluid. 
However, contrary to the Delaunay triangulation, the initial tessellation
remains intact during the entire evolution regardless of the changes of the mutual distances between the particles. 
In particular, the order of vertices  remains the same in each tetrahedron.

In the course of time a vertex of a tetrahedron can cross the opposite face. This results in the change of
the sign of $d$  indicating that the tetrahedron has turned inside out.
If two neighboring tetrahedra sharing a common face have opposite signs of $d$
then the common  face is an element of the caustic surface
and  thus becomes  a cell in the triangulation of the caustic surface. 
%The vertices of such triangles are obviously the points on the caustic. 
The caustics identified by this method are A2 singularities  according  to Arnold's classification. 
All higher order singularities are singular lines and points on caustic surface A2. 
For instance cusps  are A3 singularities on the curves  lying  on the surface A2.

In our code the  determinant $d$  is calculated for all tetrahedra only at every output time but
not at every time step  of the simulation. Therefore the numbers of the tetrahedra's  flip-flops are not available.
%which would help to isolate different caustic surfaces.
However, finding the triangle elements of caustics requires only the signs of the tetrahedra volumes.
Hence the caustics can be  obtained without recording the whole history of flip-flopping.
Instead of computing flip-flops of tetrahedra  we do it  on particles because  it is easier 
to implement numerically at every  time step  \cite{Shandarin2016}. 

Identifying  caustics  in Eulerian space is  getting complicated because they often cross each other.
If the number  of flip-flops was known for each tetrahedron then caustic triangles could be assigned 
the mean number of flip-flops of two parent tetrahedra as a proxi assisting the search of caustic surfaces.  
The caustic elements carrying the same half-integer flip-flop tag would be separated out as an individual  caustic surface. 
But we compute flip-flops only for the vertices of caustic triangles.
In order to relieve this problem we have devised two diagnostics. 

Anticipating  the number of  flip-flops to grow from external to internal  caustics 
we assign the mean number  of flip-flops computed on the vertices of the caustic triangles
$n_{\rm ff}^{\triangle} \equiv {1/3}\sum_{i=1} ^3 n_{\rm ff} (v_i)$ to the triangles 
as a diagnostic helping to isolate caustics of different generations.
We also expected that the triangles of the internal caustics must be on average smaller by size. 
We compute the length of the longest edge of the triangle $l_{\rm max} \equiv max(l_1,l_2,l_3)$ where $l_1,l_2,l_3$ 
 are the edges of a caustic triangle. Both expectations have proved to be correct as we show in section \ref{sec:CaustHigh}.
 \begin{figure} 	%	    <==================================================                    fig2
 \centering\includegraphics[width=8.cm]{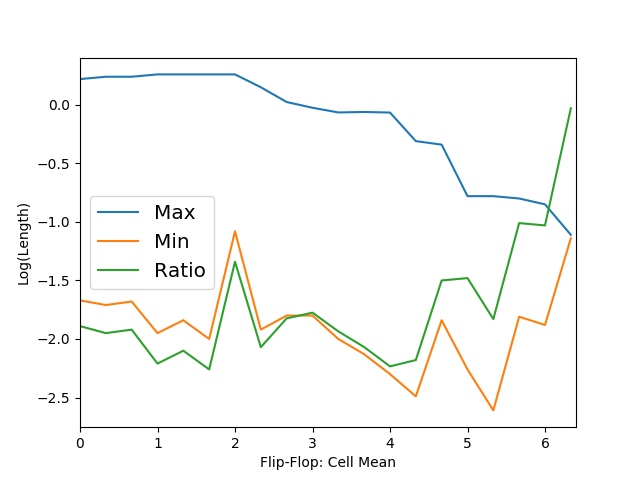}
      \caption{The  mean  of the longest $\langle l_{\rm max} \rangle$  and shortest  $\langle l_{\rm min} \rangle$ 
      edges in caustic triangles 
      as a function of $n_{\rm ff}^{\triangle}$ (Equation \ref{eq:mean_l_max}) along with
      the ratio $\langle l_{\rm min}\rangle / \langle l_{\rm max} \rangle $. }
\label{fig:lmax-ff}
\end{figure}

 We evaluate the mean  values of  $l_{\rm max}$ as a function of  $n_{\rm ff}^{\triangle}$ by averaging over all triangles 
 with the same values
 of $n_{\rm ff}^{\triangle}$
 \begin{equation}
\label{eq:mean_l_max}
  \langle{ l_{\rm max} }\rangle (n_{\rm ff}^{\triangle})  = {1 \over N_{\triangle}} \sum_{i=1}^{ N_{\triangle}}  max(l_1^{(i)}, l_2^{(i)}, l_3^{(i)}).
\end{equation}
We also  define $ \langle{ l_{\rm min} }\rangle (n_{\rm ff}^{\triangle})$ similarly to Equation  \ref{eq:mean_l_max}.
 Figure \ref{fig:lmax-ff} shows both  $\langle l_{\rm max} \rangle $  and $\langle l_{\rm min} \rangle $ along with 
 their ratio $\langle l_{\rm min}\rangle / \langle l_{\rm max} \rangle $. 
 The figure shows that  $\langle{ l_{\rm max} }\rangle (n_{\rm ff}^{\triangle})$ is monotonically 
 decreasing  from   $\langle{ l_{\rm max} }\rangle (n_{\rm ff}^{\triangle}=2)$ by more than an order of magnitude  
 at $n_{\rm ff}^{\triangle} =6.3$. 
 The caustic triangles become not only smaller but also more compact since 
 the ratio $\langle l_{\rm min}\rangle / \langle l_{\rm max} \rangle $ increased from $\sim 10^{-2}$ at $n_{\rm ff}^{\triangle} = 0$ to 
 almost $\sim 1$ at $n_{\rm ff}^{\triangle} =6.3$.
 We use the both as additional diagnostics helping to isolate independent caustic surfaces.

Our tessellation decomposes each elementary cube in five tetrahedra \citep{Shandarin2012}. 
 There are two kinds of tetrahedra in Lagrangian space: the central one 
with volume $V_c=l_0^3/3$ and four corner tetrahedra with volumes $V=l_0^3/6$ where $l_0=L/N$. Here $L$ is the size of the box
in units of $ h^{-1}$ Mpc  and $N^3$ is the number of particles in the simulation.
Initially the central tetrahedron is regular with edges  $l_c = \sqrt{2} \,l_0$.  
Four equal corner tetrahedra have three edges of length $l_1=l_0$ and three  
of length $l_2=  \sqrt{2} \,l_0$. 
	
\section{Caustics in a high sampling simulation}          %    <====================================           sec5
\label{sec:CaustHigh}

\begin{figure} 	%	 <=========================================================          fig3		
\centering\includegraphics[width=14cm]{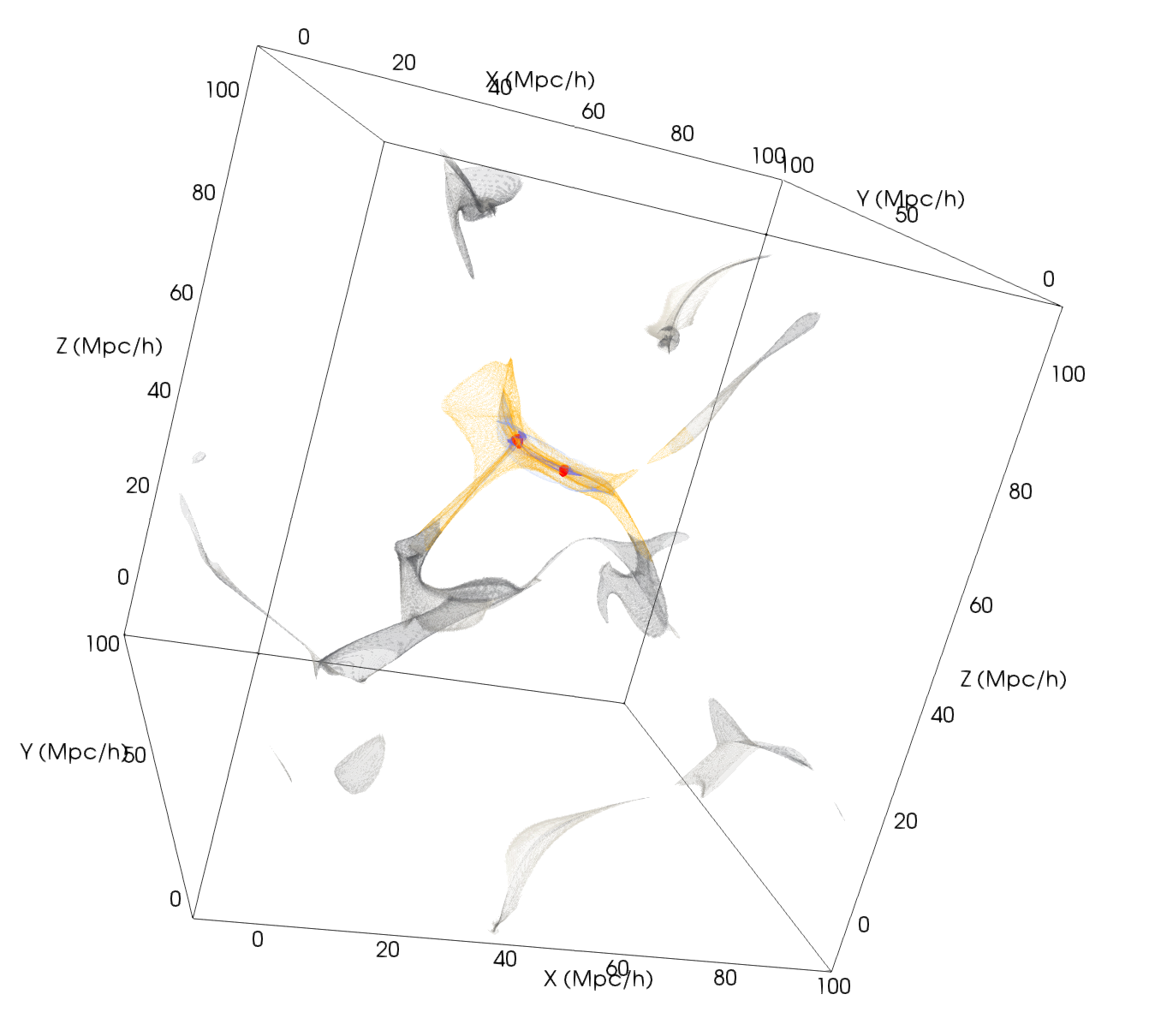}
\caption{Caustic surfaces  in Eulerian space. 
 The  colored area marks the region mostly evolved dynamically. It is selected by a spherical clip 
 with the radius of about 20 Mpc/h. }
\label{fig:full-E}
\end{figure}

We discuss only the final stage of the simulation corresponding to $a=1$ and focus on the structure that  
has reached the most advanced stage in dynamical evolution.
%The evolution of caustics with time will be discussed in the following paper analyzing a larger simulation.
The caustic surfaces in the  full simulation box are shown  in figure \ref{fig:full-E}. 
%\ref{fig:full-L} and  respectively.
There are only a few largest caustic structures with more than a thousand of caustic elements i.e. caustic triangles/cells and 
caustic vertices/particles.
However, the overall abundance of structures is not much different from  that in the HDM simulations  in \cite{2007MNRAS.380...93W}.
 
 The geometry of the external caustic shells separating 
a multi-stream regions from the single-stream region is typically simpler than that of the internal caustics.
They resembles the typical caustic structures in the ZA simulations. 
The internal caustics are much more complex and their shapes have not been systematically studied 
in cosmological N-body simulations.  Identifying and examining some of them is the major goal of our study.

The most dynamically advanced part of the caustic structure in this simulation is highlighted  by color  
in  figure  \ref{fig:full-E}.
It consists of a filament and two haloes at the both ends resembling  a dumbbell.  
It is embedded in the external  cylindrical  caustic with significantly  greater diameter than the internal filament.

\begin{figure} 	%	<=========================================================          fig4		
\centering\includegraphics[width=14cm]{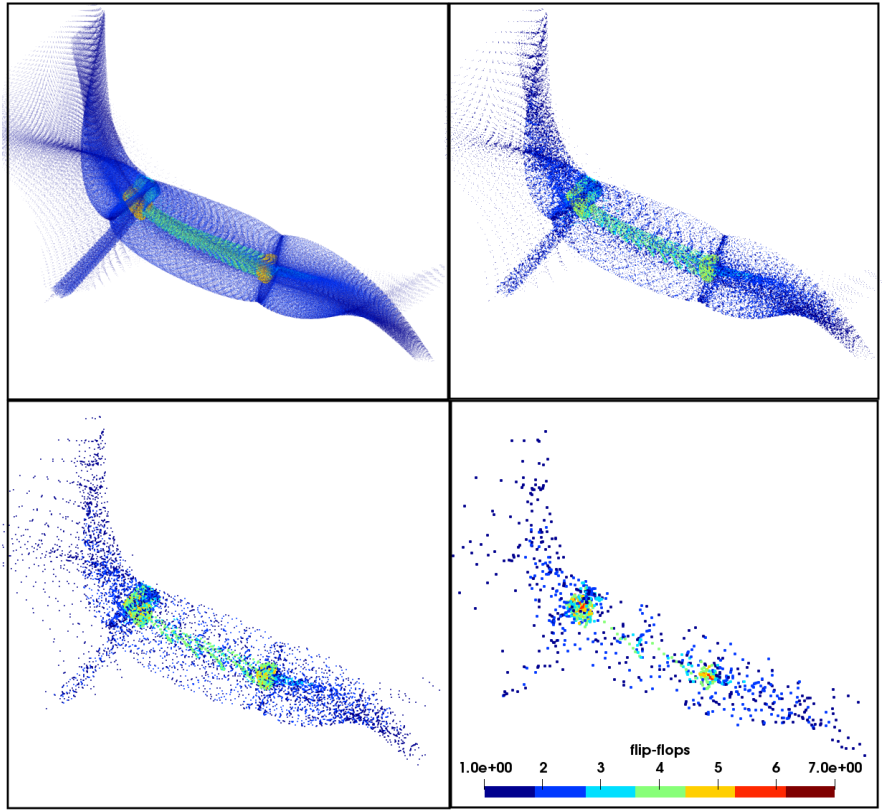}
\caption{Four images of the structure highlighted by  colour in figure \ref{fig:full-E}. It is rendered 
with different sampling densities parameterized by the ratio of the Nyquist frequency to the cutoff scale
$\kappa = k_{\rm Ny}/k_{\rm c}$.
Top left:   $\kappa = 32 $. Top right: $\kappa = 16$. Bottom left:   $\kappa = 8 $. Bottom right: $\kappa = 4$
Colour encodes the number of flip-flops passed by each particle as indicated by the legend.
Only particles that passed at least one flip-flop are shown.}
\label{fig:sampling_4pan}
\end{figure}

\subsection{Why do we need a high mass resolution simulation?}        %     <=====================                 sec5.1 
A short answer is trivial: it is  impossible  to  render or characterize a complex geometry  with insufficient 
number of elements.  
 Long time ago  is was shown in two-dimensional high-resolution simulations that the complexity of the structure 
 can be revealed only with sufficiently high mass resolution  \cite{Melott1989}.
 Using a tetrahedral tessellation of the three-dimensional manifold allows to improve  rendering the DM density with more 
 numerous  tetrahedra centers  that discloses the structure in considerably more detail  \citep{Shandarin2012,Abel2012}.
 Moreover by exploiting this technique it is possible to devise  an improved particle-mesh technique  \citep{Hahn2013,Hahn2016a}. The new technique allows to follow the evolution even in regions with very strong mixing. 
 
 Figure \ref{fig:sampling_4pan} provides a visual illustration of the importance of mass resolution by showing exactly same structure rendered with different  number of particles.
It is  useful to look at the ratio of the cutoff scale to the mass resolution scale 
$\kappa =k_{\rm Ny}/k_{\rm c}$ where $k_{\rm Ny}=(N/2)(2\pi/L)$ is the Nyquist frequency
and $k_{\rm c} = 4(2\pi/L)$ the cutoff wave number.
The top panels correspond to  $\kappa =  32$  and 16 on the left and  right side respectively
the bottom panels  to $\kappa =  8$  and 4 on the left and  right side respectively. 
In the majority of cosmological N-body simulations of the CDM universe the cutoff of initial power spectrum 
happens naturally at $\kappa \sim 1$.
The rich structure in the top-left panel ($\kappa = 32 $) is steadily fading-out as  $\kappa$ is decreasing.
Only two  weak remnants of the  haloes at the both ends of the green filament  can be identified in the 
bottom-right panel corresponding to $\kappa =4$.

It is instructive to  compare figure \ref{fig:sampling_4pan} with  the filament shown in the dot plots  of figure 2 
obtained in three simulations with a range of mass resolutions
and much smaller  force softening length  $R_{\rm s}=0.04h^{-1}$ ~Mpc  \cite{2007MNRAS.380...93W} .

\subsection{Caustics in two-dimensional slices}          %sec5.2
\begin{figure} 	%	    <================================================================= fig5
\centering\includegraphics[width=10cm]{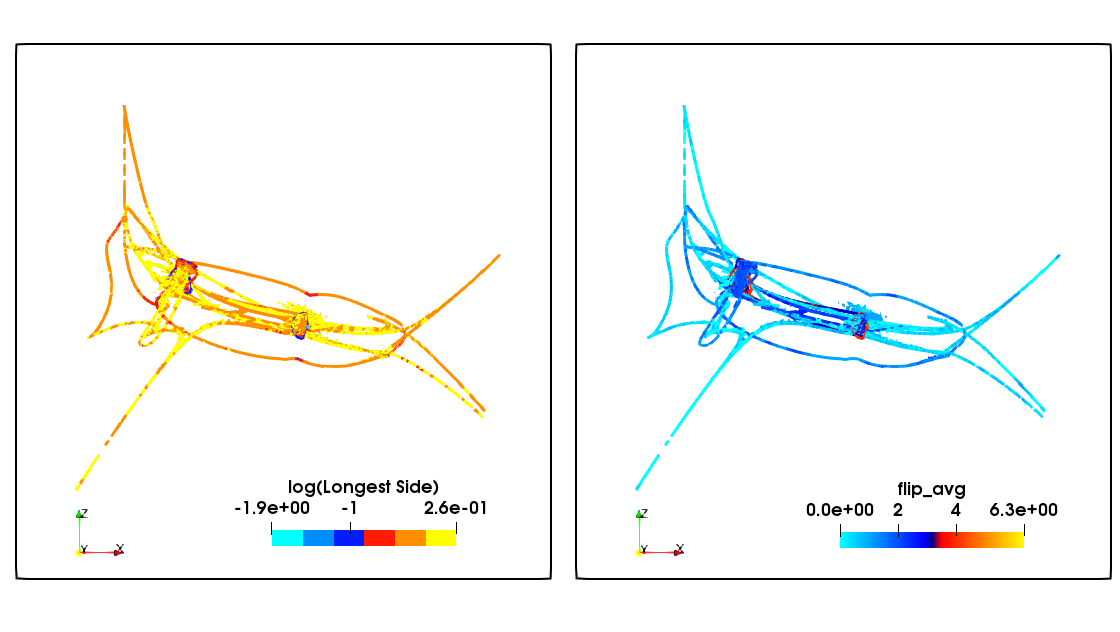} 
\centering\includegraphics[width=10cm]{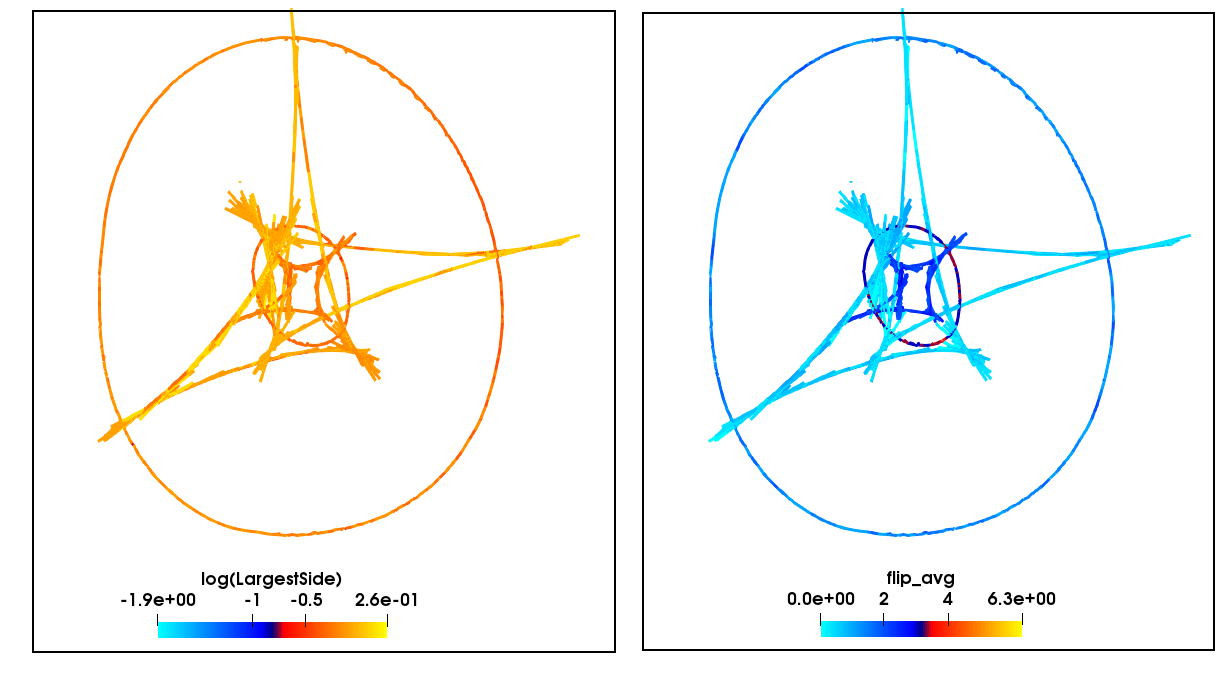}
%\centering	
\caption{Two razor thin  approximately orthogonal slices of the highlighted region in figure \ref{fig:full-E}. 
Top panels: a slice passing through two yellow halos in figure \ref{fig:sampling_4pan}. 
Bottom panels: a slice across the middle of the green filament in figure \ref{fig:sampling_4pan}. 
Colour encodes $\langle{ l_{\rm max} }\rangle$ (left panels) and $n_{\rm ff}^{\triangle}$ (right panels).
Note that the  linear scales are different in top and bottom panels.}
\label{fig:center-EL}
\end{figure}

First, we demonstrate that the structures shown in figure \ref{fig:full-E}  
represent a set of physical surfaces. Each caustic triangle is found and plotted absolutely independently
of all the rest. 
An assumption that a set of independent triangles form a continuous surfaces  inevitably predicts
that the cross-section of a plane with this set of triangles must be a discrete set  of continuous  curves. 
Figure  \ref{fig:center-EL} showing two orthogonal infinitesimal slices through the coloured caustics in figure 
\ref{fig:full-E}  unambiguously  confirms this prediction.

Two upper panels  show the same infinitesimal slice through the both red blobs in figure  \ref{fig:full-E}. 
%Colours encode $\log(l_{\rm max}/l_0)$ and $n_{\rm ff}^{\triangle}$  in the left and right panels respectively.
Two lower panels  show the slice across the structure in the middle part.
It is interesting that the two-dimensional structure 
shown in the upper panels  is   remarkably similar to figure 4 
obtained in a high resolution two-dimensional N-body simulation \citep{Melott1989} . 
This is an evidence  that  although  the number of topological types 
of caustics in 3D is greater than in 2D  nonetheless there are some similar types as in the case of the ZA.

\begin{figure} 	%	<=====================================================       fig6
\centering\includegraphics[width=12cm]{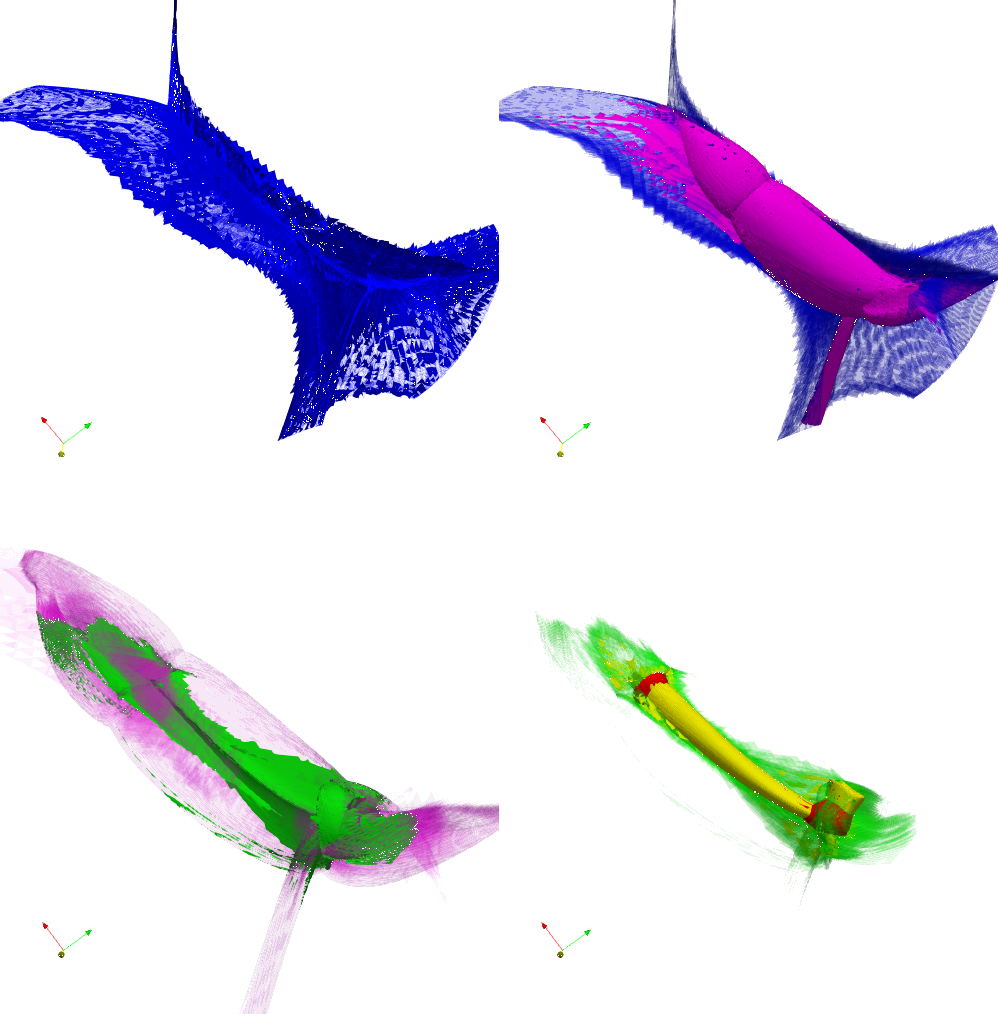} 
\caption{Four images of the caustics in the  highlighted  part of figure \ref{fig:full-E}.  
Five caustic shells are shown in blue, magenta, green, yellow and red. The surfaces in blue and magenta are 
the outermost caustics in the region.
Two lower images  have been zoomed in. The distance between two  red halo  shells seen in the lower 
right image is about 10 Mpc/h.}
%The red halo shell-like caustics are same as in figure \ref{fig:full-E}.}
\label{fig:center-E}
\end{figure}

The color bars  show the range of $\log(l_{\rm max}/l_0)$ in two panels on the left-hand side  
of figure \ref{fig:center-EL}  where $l_{\rm max}$  is the longest edge of the caustic triangles crossed by the plane.
 In the panels on the right-hand side colors show the mean number of flip-flops on the caustic triangles. 
 We are reminding that $\log( l_{\rm max}/l_0) = \log(\sqrt{2}) \approx 0.15$ and $n_{\rm ff}^{\triangle}=0$ at the initial time.
 %The color bars show the ranges of the caustic triangle sizes and the mean number of flip-flops.
  All panels demonstrate an apparent correlation of large  triangles (yellow and brown curves on the left) 
  with low counts of flip-flops  (cyan and blue curves on the right) in agreement with the claims
   in section \ref{sec:Method} and figure \ref{fig:lmax-ff}.
 Unfortunately, the correlation of small triangles  (cyan and blue lines on the left) with high counts of flip-flops
 (yellow and brown lines on the right) is not obvious. % because they are covered by large triangles in the plot.
 However, by magnifying the figure  one can see that 
 the cyan and blue curves on the left corresponding to small triangles   are covered by the yellow and brown curves 
 corresponding to large  triangles. A similar  remark can be made for small and large values of  $n_{\rm ff}^{\triangle}$
 in the right panels.
 The arrangement of the caustics in  Eulerian space -- in particular,  in the higher panels of figure  \ref{fig:center-EL} -- are 
 substantially more cumbersome than that in Lagrangian space as we discus it briefly bellow. %in Section \ref{threedim}.
Two lower panels of figure  \ref{fig:center-EL} reveal two remarkably  smooth concentric ovals.
Thus, the figure suggests that the corresponding caustics in three-dimensional space are two smooth  
approximately coaxial  oval cylinders which can be seen also in figure \ref{fig:center-E}: external in magenta and internal in yellow.  
This is a clearly non-linear feature having no analogue in the ZA. 
The both ends of the internal cylinder are compact quasi-spherical closed shells also have not seen in the ZA. 
The quasi-cylindrical caustic -- green in figure \ref{fig:sampling_4pan} --  with two  compact closed caustics 
-- red in figure \ref{fig:full-E} or yellow in figure \ref{fig:sampling_4pan} -- at the cylinder ends resemble a dumbbell.
This structure will be discussed in far greater detail later in section \ref{sec:threedim}.

\subsection{Shapes of caustics in three-dimensions}    %     <==================================      sec5.3
\label{sec:threedim}

\subsubsection{Eulerian space}
\label{eulerspace}
Now we will discuss in detail the shapes of the internal caustics in three dimensions.
The caustic that are discussed in this paper are certainly only coarse-grained approximations.
However, it is worth stressing that even coarse-grained  caustics are truly real physical objects although  
up to accuracy of the physical model.
Therefore  caustics must be  unambiguously distinguished from contour plots of the density fields because
the contour levels can be arbitrarily chosen.
A distribution of caustics   in space  represents a specific intermittent phenomenon in the sense that  it is a physical system
that has only two states: one is a discrete set of caustic surfaces and the other is empty space. 
Alternatively it can be considered as a purely
geometric structure made up from two-dimensional surfaces. Both require very specific methods of analysis.
The caustics change their positions with time but physical velocities of the particles -- the vertices of the triangulated 
surface -- cannot be considered as the velocities of the caustic because caustic surfaces evolve with phase velocities.
%We  provide visualization of several examples some of which are of peculiar shapes.
 
 As we mentioned earlier in our approach the caustic surfaces consist of mutually independent triangles 
 each of which is fully defined by two neighboring tetrahedra. 
%The triangles are identified  totally independently  of all the rest in the caustic surface 
%even some of them may belong to the same tetrahedron.
Figure \ref{fig:center-EL} suggests that setting thresholds on 
 $n_{\rm ff}^{\triangle}$ (Equation \ref{eq:mean_l_max} and the text above it) may help to isolate particular parts 
 of the total caustic surface.
As we describe in Appendix A each pattern shown in figure \ref{fig:center-E}   is  specified by  three or four 
components selected by a single  value of  $n_{\rm ff}^{\triangle}$ (Table 1 ).

The colored patch  in  figure \ref{fig:full-E}  is magnified and dissected into five distinct patterns 
displayed  in figures \ref{fig:center-E}.                         % and \ref{fig:center-L}. 
Two shells in blue and magenta shown in two upper  panels
 are the outermost caustics in the highlighted region of figure \ref{fig:full-E}.
The upper right image  shows that  the  caustics in blue and magenta   cross each other. 
In this image the blue  caustic is painted with low opacity that allows to see more of the  caustic in magenta.
The middle part of the  caustic  in magenta  looks approximately as a cylinder or tube.
The  cross section plane passing through its axis
gives an idea of its three-dimensional contour (see the lower panels in figure \ref{fig:center-EL}).
Then the caustic in magenta is plotted with low opacity in order to see the internal caustic in green
is shown in the bottom left panel of figure \ref{fig:center-E}. 
%Finally the bottom right image show a yellow filament caustic with two halo caustics in red attached to
%the filament.
 
Two red caustics in bottom right image are compact closed surfaces -- neither spherical nor ellipsoidal. 
Located exactly at the ends of the yellow tube they form a configuration made 
by two haloes connected by a cylindrical filament bounded by yellow caustic. 
We label this structure as a dumbbell. The red halos are separated by the distance of approximately 5 Mpc/h.
We are suggesting that the red caustics would be good candidates for the outermost convex caustic of the haloes
that could  be called 'splashback caustics'  on cluster scales in analogy with 'splasback radii' on galactic scales  \citep{More2015, Mansfield2017,Chang_2018}.
We provide additional arguments in section \ref{sec:dumbbell}  where we  discuss the velocity fields
as well as kinetic and potential energies in one of  them.

Very smooth shapes of  the red  and yellow caustics suggest that they are controlled
by smooth local gravitational potential which also controls the very smooth  shell in magenta. This may explain
why these components in figure \ref{fig:center-E} are so different from caustic structures known in the ZA.  
However, they are inside of green and blue shells which look more familiar
since they resemble some of the caustics predicted by   the ZA.
%One may speculate that they are closer to the relaxation in the local gravitational potential.  
They are probably formed  by the streams
falling with the velocities mostly acquired in the large-scale gravitational potential and therefore form 
the caustic structures more similar to  the caustics in the ZA.
\begin{figure} 	%	<=====================================================     fig7	
\centering\includegraphics[width=5.5cm]{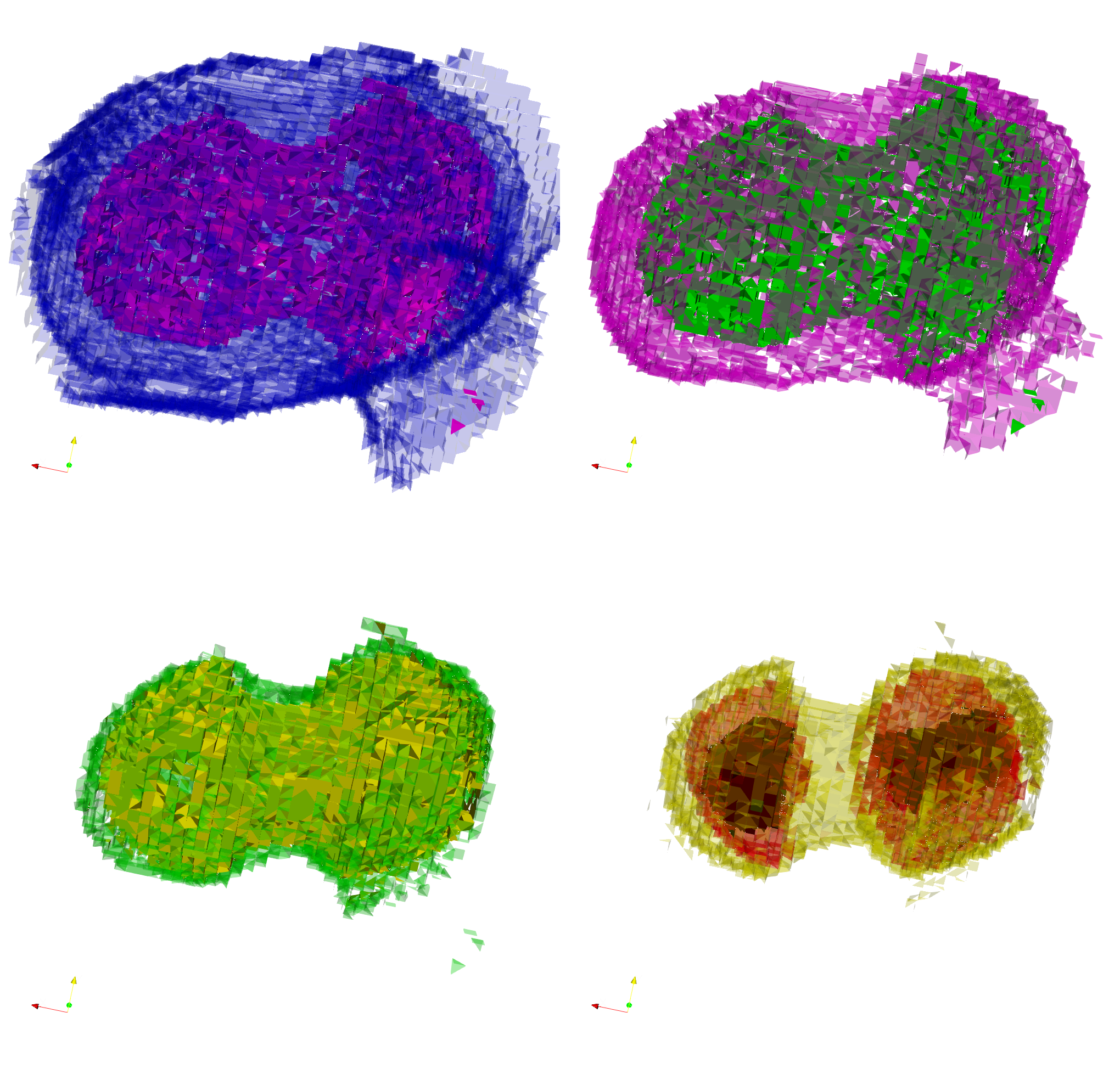}  \includegraphics[width=9.7cm]{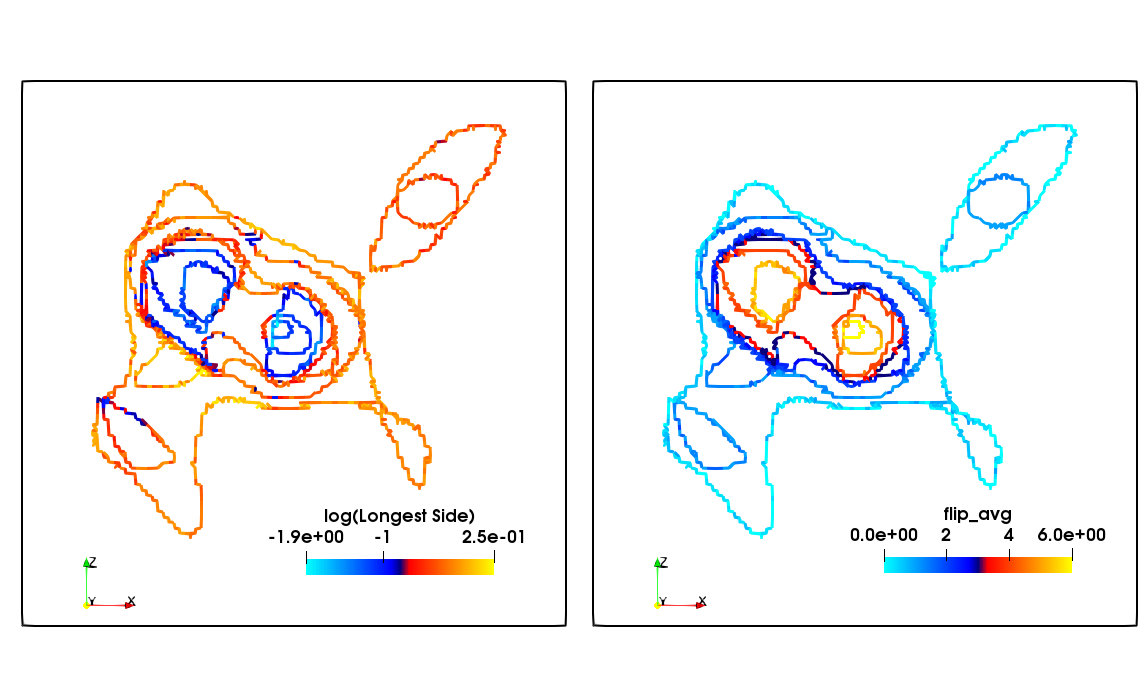}
\caption{Four images on the left are the caustics in a spherical clip with the radius about 26 Mpc/h in Lagrangian space approximately corresponding to the caustics in figure \ref{fig:center-E}.
Two right panels show a razor thin slice through the full set of caustics in Lagrangian space roughly corresponding to
the slice in the top of figure \ref{fig:center-EL}. The colour scale is the same as in figure \ref{fig:center-EL}.
Note the nesting structure of caustic surfaces in Lagrangian space.}
\label{fig:center-L}
\end{figure} 

There are small  internal caustic structure of the next generation within the red caustics.  
Their shapes are resembling the Zeldovich pancakes with thickness around 0.15 Mpc/h.
The thickness is  smaller than the force softening scale therefore their shapes may  be artefacts
of low force resolution.

\subsubsection{Lagrangian space}
\label{sec:lagrspace}
Figure \ref{fig:center-L}  shows the progenitors of five caustic shells shown in figure \ref{fig:center-E} 
in Lagrangian space. 
 In addition two black caustic shells inside the red caustics are also displayed. 
 The figure confirms the suggestion made at the end of the previous section that the caustics surfaces 
 in Lagrangian space represent the nesting structure that may be easier to disentangle than in Eulerian space.
 For instance, when two-dimensional slices in figure \ref{fig:center-EL} were discussed we mentioned that 
 the triangles with small $\langle{ l_{\rm max} }\rangle$  (left-hand panels) and 
 high  $n_{\rm ff}^{\triangle}$ (right-hand panels) could not be easily seen because
 they were obscured  by the triangles with large  $\langle{ l_{\rm max} }\rangle$  and low $n_{\rm ff}^{\triangle}$.
 But two right panels in figure \ref{fig:center-L} demonstrate the entire ranges of the both $\langle{ l_{\rm max} }\rangle$ 
 and  $n_{\rm ff}^{\triangle}$.  Thus, the internal higher generation (high  $n_{\rm ff}^{\triangle}$ -- red-yellow curves) streams  are assembled from small triangles (small $\langle{ l_{\rm max} }\rangle$ -- blue-cyan curves) in the corresponding panels of figure \ref{fig:center-L}.
 
Comparing  the caustic structure in Lagrangian space with their Eulerian counterparts one may  conclude
that it is much easier to disentangle them in  Lagrangian space than  in Eulerian space especially in dense crowded regions.
This is because the caustics  typically intersect with each other in Eulerian space but do not in Lagrangian space
where they are described by a single value function.
However, they may have common lines where $\lambda_1(\mathbf{q}) = \lambda_2(\mathbf{q})$ or 
  $\lambda_2(\mathbf{q})= \lambda_3(\mathbf{q})$  \citep{Arnold1982,Hidding2014}.

\section{The dumbbell structure}  %    <============================================          sec6
\label{sec:dumbbell}
 \begin{figure} 	%	<========================================================         fig8
\includegraphics[width=7.5cm]{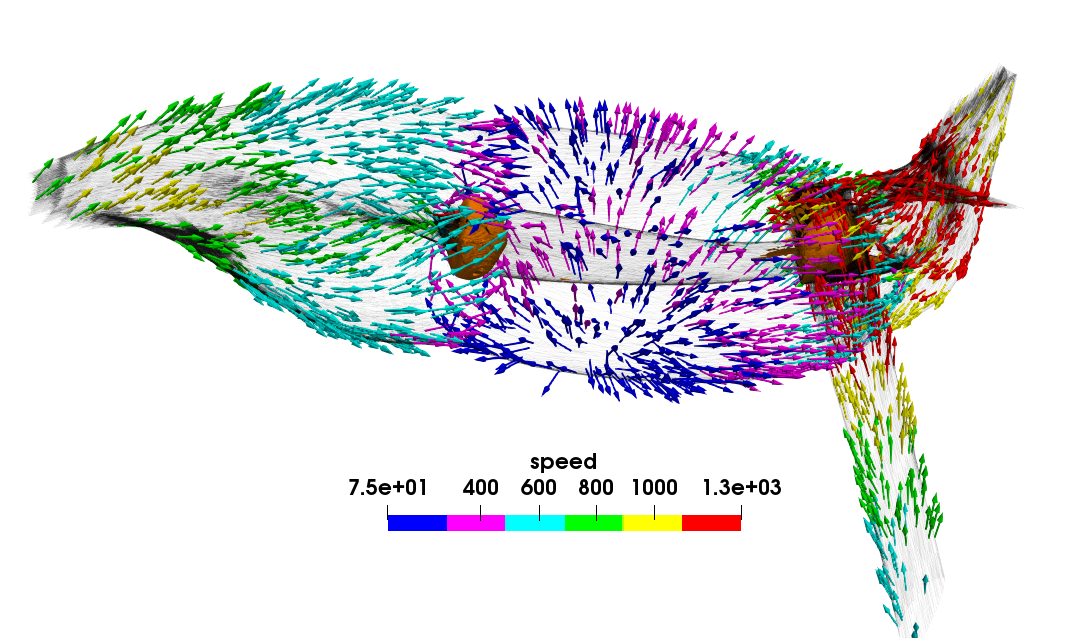}
\includegraphics[width=7.5cm]{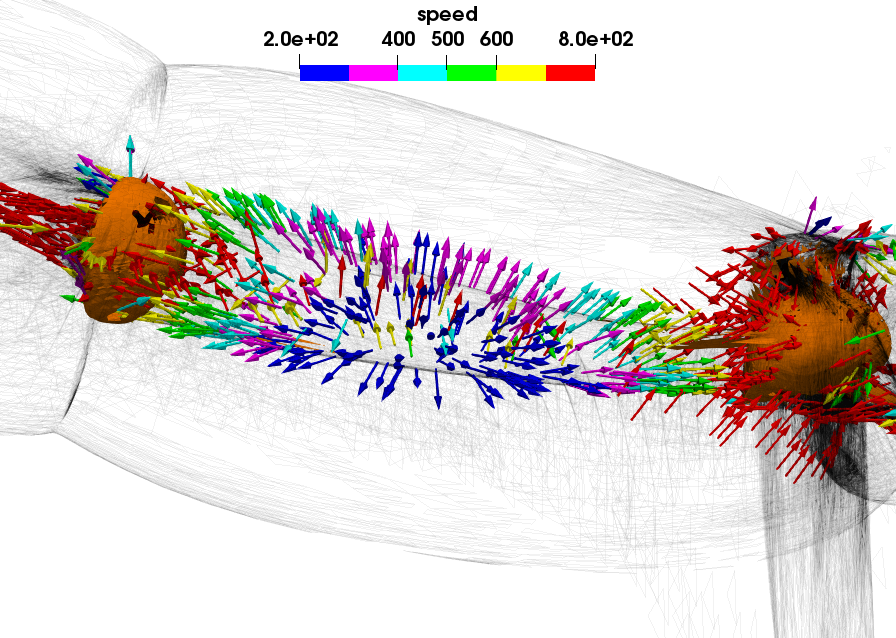}
\caption{Left panel: The velocities of the vertices of the external caustic  shell shown in figure \ref{fig:center-E} in magenta.
 Note that the arrows render only velocity directions. The speeds of particles are encoded  by colour.
 The set of arrows selected by statistically uniform spatial distribution for clarity. Two brown blobs are 
 the halo caustic shells. Brown colour allows better see the high speed particles in red.
 Right panel: The velocities of the vertices of the internal caustic  shell. Note the difference in colour scales. }
\label{fig:large-fil-vel}
\end{figure} 
\subsection{Velocity streams}	%   <=======================  subsection  6.1
 First, we consider the velocities of the particles that are  vertices of the triangle elements of two caustics
 in magenta and yellow in figure  \ref{fig:center-E} or the external and internal caustics respectively shown as
 semitransparent  gray surfaces in figure  \ref{fig:large-fil-vel}.
It is worth stressing that the velocities of the caustic vertices are not simply related to the propagation of caustic
surfaces because the latter cannot be described by  the physical velocity but requires the phase velocity.
% Two brown blobs at the ends of the internal cylinder are the caustic shells of haloes -- red in figure \ref{fig:center-E}.
  
 The velocities of particles  are shown by a set of arrows attached to a subset of the caustic vertices. 
 For clarity the subset of the vertices was  selected  according to statistically uniform spatial distribution. 
 Therefore  the appearance of the arrows does not reflect the actual density of the caustic vertices.  
 The arrows are of a constant length  and the speed is encoded by colour according to the legends in the figure.
 The caustic surfaces  in gray  are very smooth between the haloes (see also figure \ref{fig:center-EL}) 
 but outside they have rather complex connections with other parts of the caustic web.
 Those are better seen in figure \ref{fig:full-E}. 
 
 The velocity patterns are similar in both caustics. They demonstrate the steady growth of speed toward the haloes
 - the vector colours change from blue to red.  
 There is a relatively narrow  region on both caustics where the longitudinal component of the particles changes the sign.
 
 The counts of flip-flops allow a qualitative description of the trajectories of particles as a sequence of flip-flops
 that are happening  in the caustics.
 For instance, the particles that have reached the yellow caustics had experienced the first four flip-flops 
 in the following order of the caustics referred to by the colours  in figure  \ref{fig:center-E}: blue, magenta, green, and yellow.  
The slice approximately through  the middle point between the haloes in orthogonal direction is shown 
in lower panels of figure \ref{fig:center-EL}. The particles experiencing the first collapse to a caustic form
the exterior star-like configuration in cyan in the right-hand panel. 
The particles experiencing the second caustic event form the exterior oval. The third collapse corresponds to the interior
star-like configuration and the fourth collapse is happening in the interior oval. It would be naive thinking  that the particles 
have passed through the caustics shown at one instant of time because it takes a considerable time so the caustics
could evolve. Nevertheless it gives some hint about the type of caustic surfaces.
We continue the story in the next section where we analyze  the velocity structure of one of the halos. 
  
\begin{figure} 	%	<================================================                        fig9	
\centering\includegraphics[width=8cm]{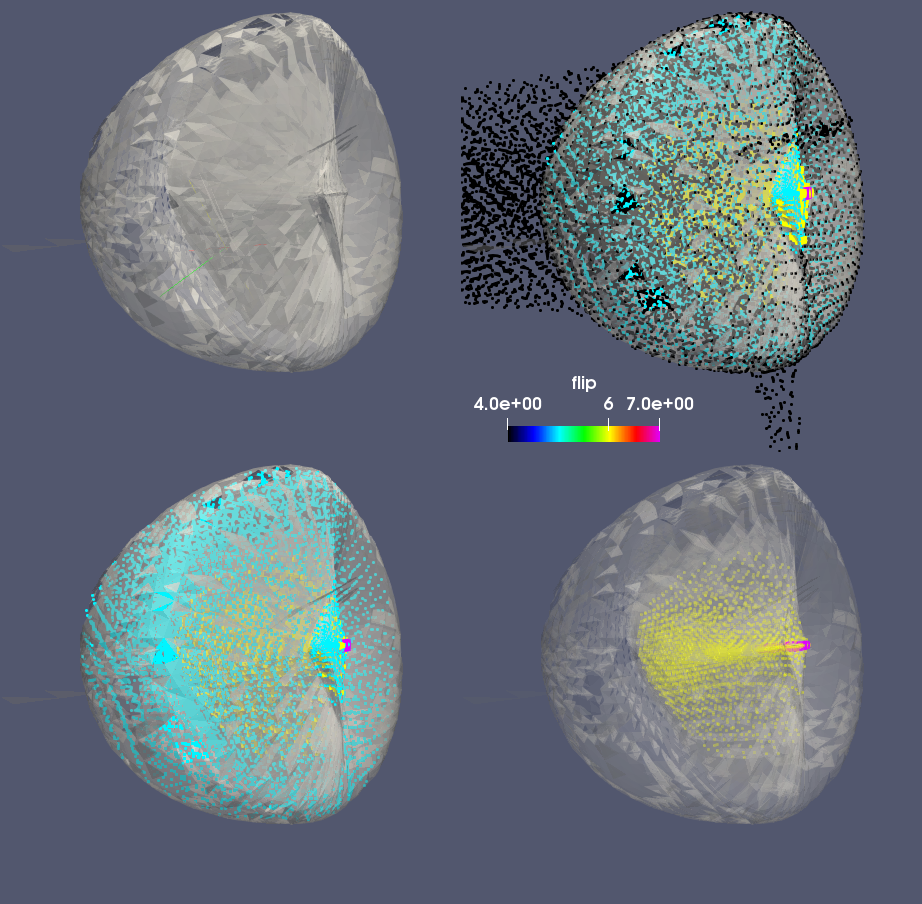}
\caption{An illustration to building the grid template of the halo identified by a caustic shell.
Top left panel: the caustic shell. Top right panel: the shell together with all the particles with flip-flop counts 
from four to seven in a small box of size 3/h Mpc. 
The colour encodes the number of flip-flops: black - 4, cyan - 5, yellow - 6, and red - 7 flip-flops.
Bottom left panel: the shell and the particles with  flip-flop counts from five to seven. 
Bottom right panel: the shell and the particles with  flip-flop counts from six to seven.} 
\label{fig:ff-4-7}
\end{figure}   
%The caustic shell and the particles with  flip-flop counts from four to seven within a small box of size 
%Note the difference in the orientations in figure \ref{fig:large-fil-vel}
%and figure \ref{fig:ff-4-7} 
\subsection{The  halo bounded by a caustic}   %          <=========================                         subsection  6.2
Here we consider the  structure of the  streams inside one of two brown caustic shells shown in figure \ref{fig:large-fil-vel}.
The shell is in the left in both images. The caustic is closed and 
convex however, it's shape neither spherical nor ellipsoidal. Therefore we will begin with an explanation how 
the particles in all streams were found within the caustic shell. 

\subsubsection{Building a mask for asymmetrical halo}   % <===============  subsection   6.2.1 
In order to speed up computing  we begin with selecting all the particles in a small cubic subbox of size 3 Mpc/h  
that completely  encloses the caustic boundary.   
The box contains 79563 particles with the total mass $3.9\times10^{14} M_{\bigodot}/h$.  
 Figure \ref{fig:ff-4-7} shows the caustic shell in top-left panel.  The image in top-right panel
 shows the   caustic surface and all particles in the box with  $n_{\rm ff} \ge 4$. 
 Some  particles coloured in black  with  $n_{\rm ff} = 4$ are outside of the caustic shell but as we will see later the most of such particles are inside.
 Two images in the bottom panels show the caustic shell
 and all the particles in the box with $n_{\rm ff} \ge 5$ and $n_{\rm ff} \ge 6 $ on the left and right of the figure respectively.
 Therefore the particles with  $n_{\rm ff} \ge 5$  can be used for building the mask that later may be used  for selecting 
 all particles  inside the caustic shell. 
 In our case we map the particles with $n_{\rm ff} \ge 5$ into an auxiliary cubic grid using 
the nearest-grid-point (NGP) scheme to label the grid points inside the caustic shell. The labeled grid points
make the geometrical template for the halo based only on physics with no additional assumptions or parameters.
The last step is to identify all particles in the subbox that are near the labeled grid points.

\begin{figure} 	%	<=====================================================           fig10
\includegraphics[width=8cm]{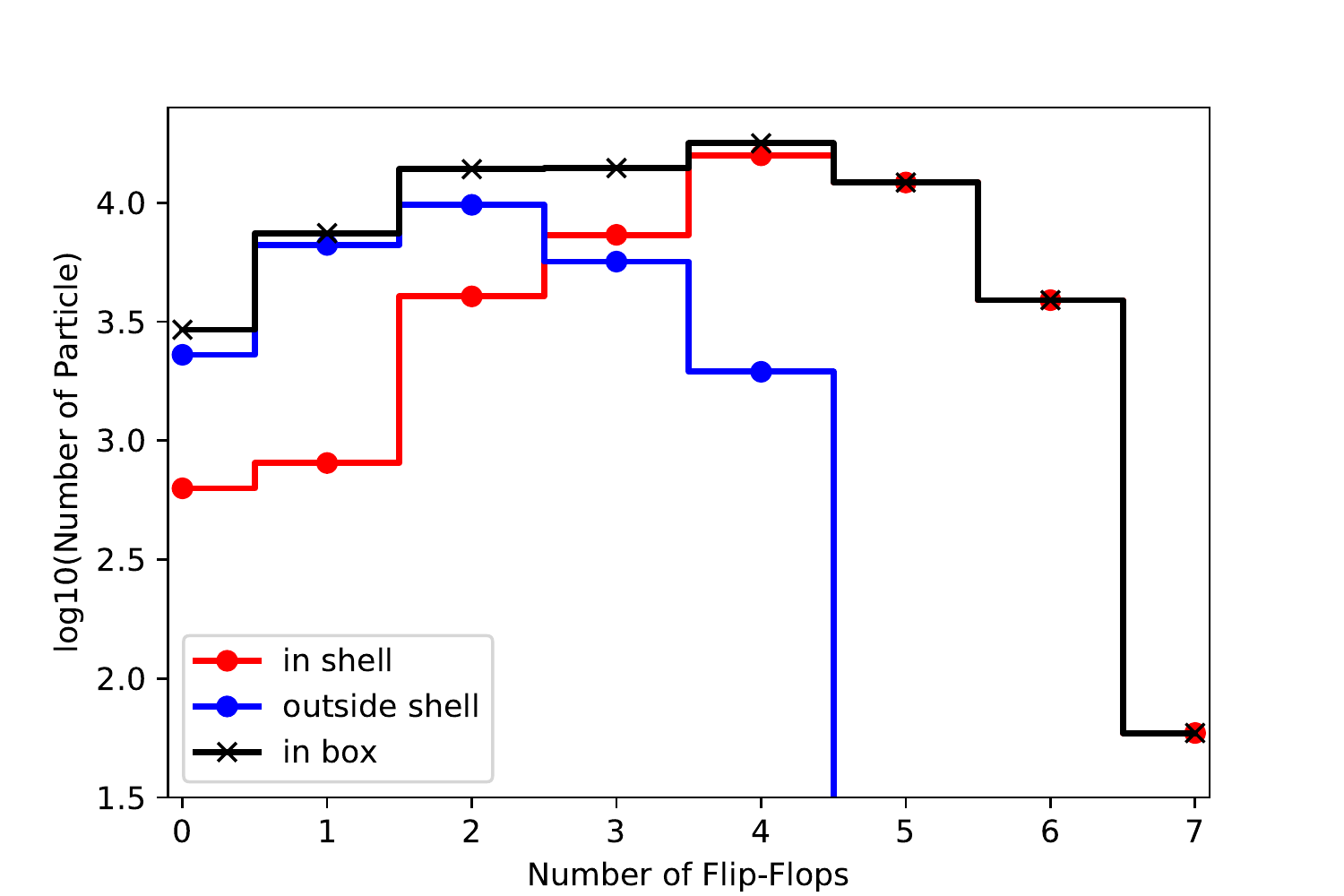}
\includegraphics[width=8cm]{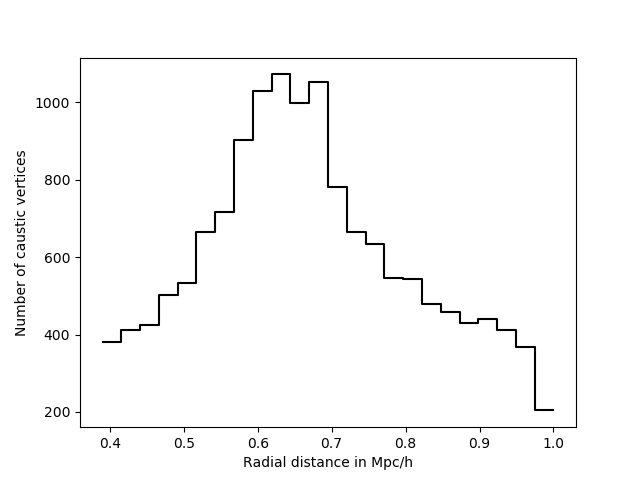}
\caption{Left panel histograms:  the number of particles with different counts of flip-flops. The labels with the colour of each 
histogram indicate  how the particles were selected.
Right panel histogram: the number of caustic vertices vs. the distance from the center of the caustic shell.}
\label{fig:shell-box-parts}
\end{figure}

The red, blue and black histograms on the left of figure \ref{fig:shell-box-parts} show the number of particles 
inside and outside of the caustic boundary as well as the total in the subbox  respectively.
The blue histogram  clearly shows that there are no particles outside the shell with $n_{\rm ff} \ge 5$
and the most of particles with $n_{\rm ff} = 4$  are inside the shell. 
The red histogram shows that the number of particles with $ n_{\rm ff} = 4 $ and  $ n_{\rm ff} = 3 $ 
inside the shell is greater than that outside the shell however, if $ n_{\rm ff} < 3 $ the opposite is correct.
The distribution of the distances of the caustic vertices from the center of mass of the caustic particles 
shown on the  right of figure \ref{fig:shell-box-parts} provides some notion of its asymmetry.

 The total volume of the template region is $V_{\rm templ} =$ 2.4 (Mpc/h)$^3$ which is a good approximation of the volume
 within the caustic shell. 
  
The counts of particles with 7, 6, 5 , 4 flip-flops within the caustic shell are 59,  3893, 12158, and 15817 respectively. 
The total  number is 31927 making the mass within the shell $M(n_{\rm ff} \ge 4) = 1.6 \times 10^{14} M_{\bigodot}/h$.  
However, the shell contains also streams with 0, 1, 2, and 3 flip-flops with 630, 805, 4037, and 7325 particles respectively.
Thus the total number of all particles  within the caustic shell is 44724  and the mass becomes 
 $M(0 \le{n_{\rm ff}}\le 7) \approx 2.2 \times 10^{14} M_{\bigodot}/h$. 
Dividing the mass by the volume within the caustic shell we estimate the mean density within the shell  as 
$9.2 \times 10^{13}M_{\bigodot}$h$^2$/Mpc$^3$ which is about  thousand times greater than the mean mass density
in the universe.
%3.17e14/2.4 = 132 083 333 333 333.34 = 1.32e14 M_s/Mpc^3
\begin{figure} 	%	<======================================================        fig11	
\centering\includegraphics[width=8cm]{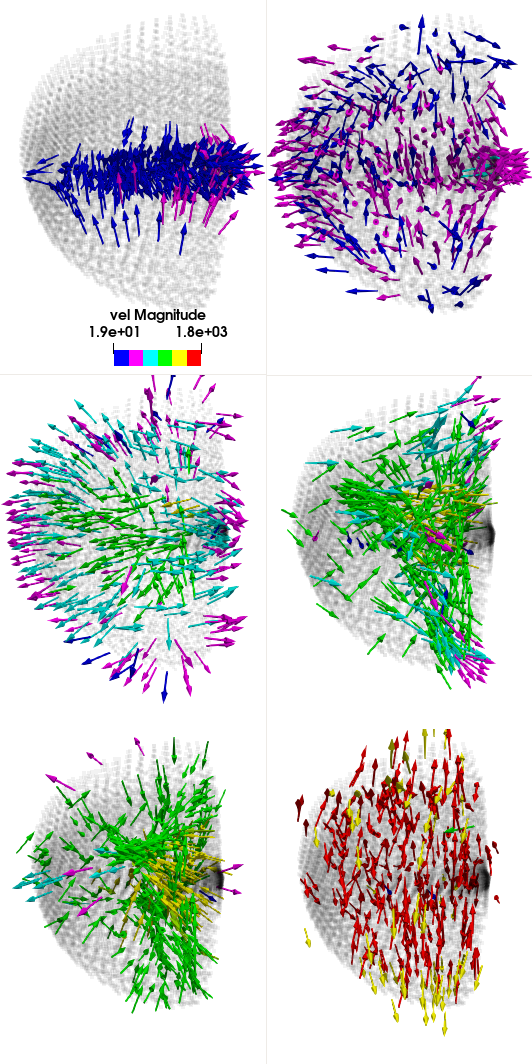}
\caption{Velocities in the streams selected by the flip-flop count  $n_{\rm ff}$. 
Randomly selected subsamples of the particles within the caustic shell are shown in six panels. 
Top-left panel : $n_{\rm ff} = 7~ {\rm or} ~6$. Top-right panel:  $n_{\rm ff} = 5$. 
Middle-left panel : $n_{\rm ff} = 4$. Middle-right panel:  $n_{\rm ff} = 3$.
Bottom-left panel : $n_{\rm ff} = 2$. Bottom-right panel:  $n_{\rm ff} = 1~ {\rm or} ~0$.
The caustic is shown by semi transparent gray surface.  
Vectors show only the direction of the velocities. 
Colour encodes the ranges of speed in km/s:  19 $<$ blue $<$ 316 $<$ magenta $<$ 613 $<$ cyan $<$ 910 $<$ green $<$ 1206 $<$ yellow $<$ 1503 $<$ red $<$ 1800.  }
\label{fig:ff-0-7IN}
\end{figure} 
 
 \subsection{Velocity field within the caustic shell}       %            <========================       Subsection  6.3
 \label{subs: subs6-3}
 Figure \ref{fig:ff-0-7IN} illustrates the velocity fields in the streams selected by the values of flip-flops. 
 For the purpose of clarity all  arrows have the same length thus showing only  the directions of the velocities 
 of a randomly selected particle set for every stream.  
 The colour encodes the speed.
  The discrete set of colors represents equally spaced intervals
 in the range from the minimum to maximum  of the particle speeds as described in the caption to figure \ref{fig:ff-0-7IN}.  
The caption of the figure provides the description of the images.
 %Two top panels show streams with seven and six flip-flops on the left and five flip-flops on the right hand side.  Two panels in the middle
%display streams with four  and three flip-flops and two panels in the bottom exhibit streams with two flip-flops  in the left  and
%with one and zero flip-flops in the right hand side.

The velocity streams seems to form the patterns of four types. 
\begin{enumerate}
\item The slowest particles with 7 and 6 flip-flops - blue in top-left panel show the infall
of particles on the pancake-like caustic structure. 
\item The particles  with 5 and 4 flip-flops shown in top-right  and left-middle panels  seem to relate mostly
 to the caustic shell. The particles in magenta and blue  indicate
 that the speed decreases when the particles approach the caustic boundary from inside. 
\item Two panels dominated by green arrows (right-middle and left-bottom panels) 
seem to fall into the central part and building up another caustic structure with different geometry than the caustic shell. 
Compare with the images in bottom panels of figure  \ref{fig:center-EL}.
%Compare with the structures in the center of the images in the middle
%row of Fig. \ref{fig:center-EL}.
\item  Finally the fastest particles (yellow and red on the right side of the bottom row) seem to  zoom
through the caustic shell practically  ignoring its gravitational field.
\end{enumerate}

 \begin{figure} 	%	                     <======================================================           fig12
\centering\includegraphics[width=10cm]{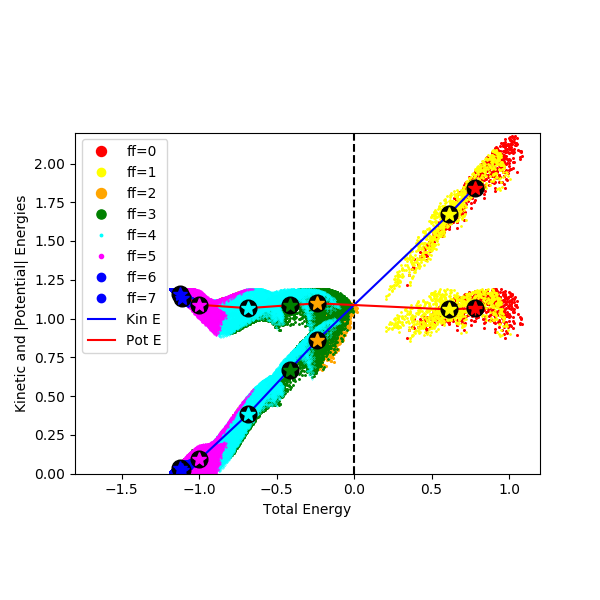}
\caption{Kinetic and  negative potential energies  of the particles in the halo  are plotted  as the functions of the total energy
shown on the horizontal in units of $10^{59}$ ergs/h. The streams are indicated by different colours. The coloured dots in the legend are
proportional to the sizes of the points in the plots. Stars in black circles show the corresponding mean values for each stream.
The blue  and red lines connecting the stars show the kinetic and potential energies respectively. 
The vertical dashed line marks zero of the total energy.  The mean energies of the stream with $n_{\rm ff} = 3$ (in green)
approximately satisfy the virial ratio.}
\label{fig:halo-part-energ}
\end{figure} 

 \subsection{The energy of the halo}       %   <=======================  Subsection 6.4
\subsubsection{The energy distribution in the halo}   %  <=================================     subsection   6.4.1
The distribution of particle energies in the halo provides an additional diagnostic of its dynamical structure.
We evaluate  kinetic,  potential and the total energies for all particles in the halo:
$K_{\rm i}=0.5 m_p v_i^2 $, $U_{\rm i}= -G m_p^2\Sigma_{j=1}^{N_{p}} |{\bf r}_i -{\bf r}_j|^{-1} $  (${\rm j } \not={\rm i}$)
and $E_{\rm i} =K_{\rm i} +U_{\rm i}$ where ${\rm i}$ is the numerical label of a particle.
In order to be consistent with the simulation we used a similar  condition for softening gravity: 
 in cases when the separation of two particles
 $|{\bf r}_i -{\bf r}_j|$ is less than the force softening scale $R_{\rm s}=0.8h^{-1}$ ~Mpc  it was set to $R_{\rm s}$.
Figure \ref{fig:halo-part-energ} shows the dot plots of the kinetic $K_{\rm i}$ and negative potential $-U_{\rm i}$ energies 
as functions of total $E_{\rm i}$ energy.  
The particles in each stream have its own colour. The colour scheme is the same as in figure \ref{fig:ff-0-7IN}. 
The  mean potential energy is almost identical in every stream. 
Therefore the relation between the mean total and  kinetic energies is almost exactly linear
$\langle K \rangle \approx 1.1 + \langle E \rangle$ in units of $10^{59}$ ergs/h.

The mean total energy   is steadily decreasing   with the grows of the number of flip-flops in the stream. 
It is negative for streams with $n_{\rm ff} \geq 2$
while $E$  is  positive   for the streams with $n_{\rm ff}< 2$. The virial ratio is
$\langle K\rangle / \langle|U|\rangle \approx 1.6$ and $ 1.7$ for streams with $n_{\rm ff}=1$ and $0$ respectively.

Figure  \ref{fig:halo-part-energ}  is in agreement with the last statement of the previous subsection.
The streams with  $n_{\rm ff} \le 1$   are unlikely gravitationally bound to  the halo. Therefore their input
into the kinetic energy of the halo can be excluded from the total energy, but the input to the potential energy 
probably should be kept because it is not affected by their speeds.

 \subsubsection{The total energy inside the shell}   %  <=================================     subsection   6.4.2
Direct summation of kinetic energies of all particles results in $K =  \Sigma_{\rm i}K_{\rm i} = 1.27\times 10^{63}$ ergs/h 
and summation
of potential energies over all pairs of particles gives   $U = \Sigma_{\rm i} U_{\rm i} = -2.4\times 10^{63}$ ergs.
The ratio $K/|K+U| = 1.12$ that is 12\% higher than the perfect virial ratio. 
The fraction of kinetic energy  by two fastest streams with $n_{\rm ff} \le 1$ is about 13\%. 
Thus if it is excluded from the total energy balance then  the ratio become $K/|K+U| = 0.87$ which is 13\%
less than exact virial ratio. 
Both seem to be in a reasonable agreement  with the virial ratio if one takes 
into account that the both $K$ and $U$ are supposed to be averaged over time  and the dynamical system is assumed 
to be stable.
Neither requirement is exactly fulfilled in this case.
\begin{figure} 	%	              <========================================================         fig13
\includegraphics[width=8cm]{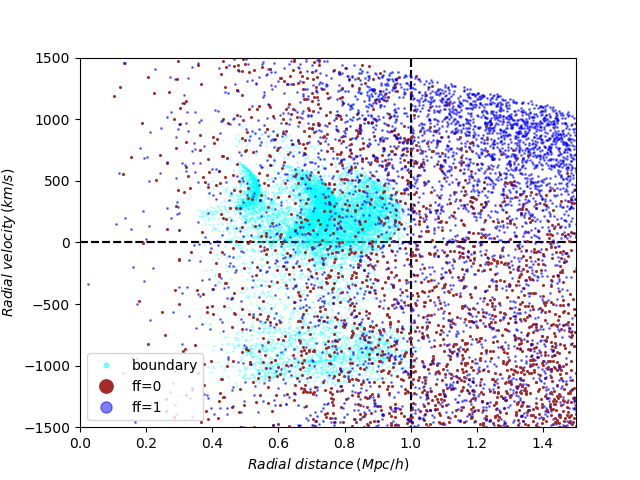}
\includegraphics[width=8cm]{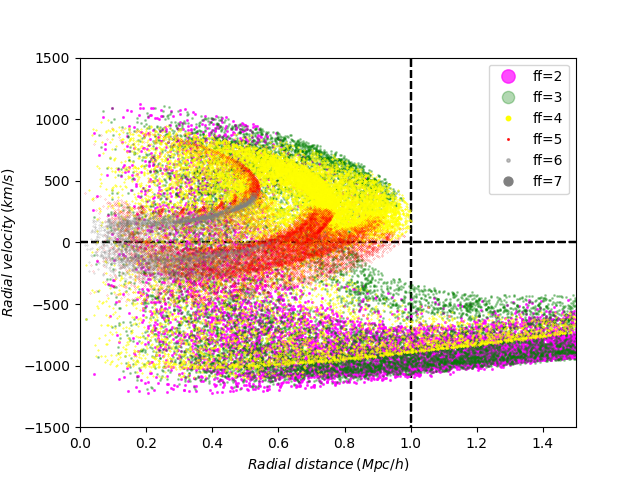}
%\centering\includegraphics[width=8.7cm]{halo-phase-space-4567-h27.png}
\caption{Phase space structure of the halo bounded by the caustic shell. 
Left-hand panel:  two streams with $n_{\rm ff} \le 1$  (brown and blue)  and the boundary particles in cyan. 
These are the fastest streams in the halo. The gravity of the halo does not  noticeably affect these  streams. 
Right-hand  panel: five streams with $n_{\rm ff}\ge 2$. Colour encodes the number of flip-flops
in every stream.
The vertical dashed line indicates the most distant elements of the caustic shell from the halo center
 of mass. The horizontal  dashed line separates inflow and outflow particles.
The sizes of the dots in the the colour legend are proportional though not equal  to the sizes of the dots in the plots.    }
\label{fig:phase-sp-4567}
\end{figure} 
 \subsection{The halo phase space}       %  <===============================       Subsection 6.5
Unfortunately it is not feasible to illustrate of the velocity field in the caustic shell region 
in the full six-dimensional  phase space.
Therefore figure \ref{fig:phase-sp-4567}  presents a commonly used two-dimensional scatter plots of the radial velocity 
 vs.  radial distance from the center of mass of the halo. 
The radial velocity is measured  with respect to the mean velocity of the halo. 
It is worth emphasizing that the halo is at the origination stage therefore its outermost  caustic
has a rather irregular shape.

The panel on the left of figure \ref{fig:phase-sp-4567} shows two coloured streams (brown and blue) with $n_{\rm ff} \le 1$ 
as well as the caustic particles in cyan. 
The range of the radial velocities of the boundary caustic vertices is also quite large.
The distribution of particles  seems to show little influence of the halo gravity on these streams.
 For the  particles with zero flip-flops (brown) this is the first run inside a multi-stream region after they 
 crossed the blue caustic from outside of the multi-stream region without experiencing a flip-flop.  
When they reach the caustic between the three-stream flow 
from the inside of the three-stream region they experience the first flip-flop and return back to the three-stream flow region. Their color changes from brown to blue in the phase space plot. The next metamorphose of some of them 
happens when they reach from inside the caustic in magenta.
However, no particle with $n_{\rm ff} \le 1$ that enter the halo is gravitationally bound (see figure \ref{fig:halo-part-energ})
to the halo and therefore all of them exit the halo.

The panel on the right-hand side of the figure shows the phase space structure of the streams 
that become gravitationally bound in the halo.
There are three types of particles entering the halo. All of them experienced a flip-flop event
in the caustic in magenta.  Then some of them directly enter the halo while 
another group experiences the second  flip-flop event in the green caustic and then enter the halo. 
The third set of particles also experiences a caustic metamorphose in the yellow caustic and only then enter the halo.
There are three streams with $2 \le n_{\rm ff}  \le 4$ entering the halo and no particles with $n_{\rm ff}  \ge 2$ 
leave the halo. 
Since the caustic metamorphoses results in the growth of flip-flops  therefore the particles in magenta
have experienced two flip-flops, green particles three and yellow particles four flip-flops. The red particles
have experienced the fifth flip-flop inside the halo and some of them have became black  experiencing
the sixth and seventh flip-flops.

The phase space pattern shown on the right-hand side of the figure  \ref{fig:phase-sp-4567} is typical for haloes. 
The external streams with  $2 \le n_{\rm ff}  \le 4$             %on the right from the vertical dashed line
are entering the halo with negative radial velocities.  %  by crossing the vertical dashed line.
They fall on the central region and after passing it  the particles get going away with  positive radial velocities.  
This results in the instantaneous leaps  of the particles from the lower part onto  the upper part in the phase space diagram. The positive radial velocities of the particles are gradually decreasing  with the growth  of the radial distances.
 At  some distance the fluid elements experience another  flip-flop resulting in the formation of the caustic 
 boundary of the halo.  
 Comparing this figure with figure \ref{fig:center-E} one can anticipate  strong
anisotropies of the streams entering  the halo.
%However, the caustics referred only by the colours because the literal 
 
   It is worth stressing that the pattern  of the phase space shown in figure \ref{fig:phase-sp-4567}  
   is strongly affected by the lack of spherical symmetry in the caustic boundary of the halo. 
   It is highlighted by the cyan particles in the left-hand panel of the figure. 
   The caustic boundary operates as a  splashback shell described in  \citep{Mansfield2017}
  who stressed that  "the splashback shells are generally highly aspherical, with non-ellipsoidal oval shapes 
  being particularly common".

\section{Summary}  %    <========================================================             sec7
\label{sec:summary}
The most common approach to the study of the DM structures consists in the analysis of the DM density 
and velocities in cosmological N-body simulations. 
We are suggesting a complimentary technique based on identifying and dissecting the  caustic structure.
%First, it worth emphasizing   
Caustics are physical objects while the isodensity contours are mathematical constructions. 
This is  the major difference between caustics and the  density contours  in DM medium.
Both can be used for the description of the structures formed in the non-liner regime.
However, the number and values of density levels can be chosen arbitrary that would result in 
certain arbitrariness of their positions in space and geometrical shapes.
In contrast to the density contours both the number of the caustic surfaces, their positions in space and shapes 
are completely fixed by physics.   
Therefore the characterization of the DM structures can be done uniquely and unambiguously by locating caustics in 
cosmological N-body simulations and studying their shapes.

The caustics in our simulation can not be associated with fine grained phase space streams.
However the estimate of the DM density from N-body simulations is also only a coarse-grained field. 
Thus the caustics in N-body simulations should be considered as the coarse-grained  approximation obtained 
from a coarse-grained phase space.
These results can not be directly used for the estimates of the density in the fine-grained caustics.
On the other hand we show that the coarse-grained caustics can provide reasonably accurate
 boundaries for the haloes and  filaments.  
Our major findings are summarized bellow.
\begin{enumerate}
\item We have identified the caustic structure resembling a dumbbell.
It consists of two halo shells connected by a quasi-cylinder caustic. 
Many examples of this structure can be seen in the density plots in \cite{Abel2012, Kaehler-etal12, More2015}.
Often  two or three dumbbells sharing the halo between two filaments are arranged in a linear pattern.
In our idealized simulation the dumbbell structure is coaxially embedded into another quasi-cylindrical caustic
with roughly three times greater diameter. The velocity patterns in the both cylindrical caustics  suggest
that the particles move toward the haloes with increasing speed.
This is an agreement with the prediction made long ago, see e.g. \cite{Shandarin1989}. 
Finding this structure in a small simulation suggests that it undoubtedly must be ubiquitous.   

\item We introduce a novel method of finding DM haloes based on the assumption  that the outermost
convex caustic is a natural definition  of a physical halo boundary. This is similar to the suggestion of 
\cite{More2015,Mansfield2017}. However, our approach has a significant difference.  
We build caustic surfaces and find the outermost convex caustics.
In the course of evolution each particle accumulates the count of flip-flops which is used for identifying
different streams \cite{Shandarin2014}. 
This parameter is also used as  additional control of building the caustics since they are boundaries
between regions with different number of flip-flops in Lagrangian space. We also used the flip-flop counts
in order to identify the streams that are fully within the caustic bound. The particles of such streams were used
for building a mask (see figure \ref{fig:shell-box-parts}) for the halo. The mask is used for finding all the particles from
every stream exactly inside of an asymmetric boundary of the halo. The entire procedure  is utterly free of other 
assumptions  about the shape of the boundary.  There is also no need   in   fitting parameters.

\item Knowing  all particles inside the halo boundary it was easy to evaluate the mass and 
kinetic, potential and total energies of the halo: $M(0 \le{n_{\rm ff}}\le 7)= 2.2 \times 10^{14} M_{\bigodot}$,
$K(0 \le{n_{\rm ff}}\le 7)  = 1.27\times 10^{63}$ ergs, 
$U(0 \le{n_{\rm ff}}\le 7)  = -2.4\times 10^{63}$ ergs and 
$E = K+U = - 1.13\times 10^{63}$ ergs. The ratio $K/|E| = 1.12$ is about 12\% higher than the virial value.

It is  remarkable that the mean potential energies of the particles in separate streams are very similar
while the mean kinetic energy of the particles  decreases steadily with increasing $n_{\rm ff}$.
The mean kinetic energy of the streams with $n_{\rm ff} \le 1$  is about 65\% higher of absolute value of
the corresponding mean potential energy. 
Therefore it is likely that these streams are not gravitationally bound to the halo. 
This conclusion is in excellent agreement
with the flow patterns of these streams  in  figure  \ref{fig:ff-0-7IN} and \ref{fig:phase-sp-4567}.
%\item gravitationally unbound streams

\item The two-dimensional phase space of the halo shows that all streams with $2 \le n_{\rm ff} \le 4 $ cross
the boundary caustic of the halo with negative velocities with respect to the mean velocity of the halo.
No particles with $ n_{\rm ff} \ge 2  $ exit from the halo. The growth of $n_{\rm ff}$  inside the caustic boundary
is the evidence that the halo is gravitationally bound.
 
%\item Lagrangian space
\end{enumerate}

\acknowledgments
I thank  Nesar Ramachandra for helping to run the N-body simulation and useful discussions.
The author acknowledges the support from DOE BES Award DE-SC0019474. 

\bibliographystyle{JHEP}
\bibliography{library}
%\bibliography{refs_for_caustic_papers}
\begin{verbatim}

\end{verbatim}
\newpage

\appendix
\section{Identifying distinct caustic structures}
A caustic element -- a caustic triangle and its vertices -- is defined as the shared face of a pair of neighboring tetrahedra 
having opposite signs of  volumes evaluated by Equation \ref{eq:det}. 
Each caustic triangle is treated as an independent entity. 
Therefore the caustic surfaces can be completely determined by a local condition. 
The mean value of the flip-flop counts on three vertices of a caustic triangle  
is discrete:  $n_{\rm ff}^{\triangle} =n \times (1/3)$ (n -- integer) because
the  number of counts of flip-flops on each vertex is integer.
The number of the caustic elements for each value of  $n_{\rm ff}^{\triangle}$  
is given in Table 1. The table also provides the reference to the colour of five caustics
shown  in figure \ref{fig:center-EL}. Two innermost caustics can be seen only in Lagrangian space 
shown as  black blob  inside red caustic shells in figure \ref{fig:center-L}.

Table 1 shows that  the most of triangles with $n_{\rm ff}^{\triangle} =1$ are in the caustic in magenta but about
10\% of them belong to the blue caustic. Similarly the majority of triangles with $n_{\rm ff}^{\triangle} =2$ are in
the green caustic but about 10\%  belong the the caustic in magenta.
Probably it is caused by inconsistency between counting flip-flops on vertices of the tessellation tetrahedra 
and identifying the caustic triangles by comparing parities of the tetrahedra themselves.
We leave solving this problem for the further work.

\begin{table}{ Table 1. The number of caustic triangles N($\Delta)$ for each $n_{\rm ff}^{\triangle}$.}\\[1ex]
\begin{tabular} {l l r r r r r}
\hline
                 & $n_{\rm ff}^{\triangle}$  &0		&1/3		&2/3	   	&1	& Total \\
Blue     &N($\Delta)$	    &28439  	&41737		&27141	  	&5195	&78112\\
 \hline
   & $n_{\rm ff}^{\triangle}$	    &1		&4/3		&5/3		&2	& Total	 \\
Magenta     &N($\Delta)$	                &51518     	&33484  	&21954		&3057	&110013\\
\hline
& $n_{\rm ff}^{\triangle}$ 	    &2		&7/3		&8/3 		&	&Total	 \\
 Green  &N($\Delta)$                     &49245	&23261 	&14600		&	&87106\\
\hline
& $n_{\rm ff}^{\triangle}$ 	    &3		&10/3		&11/3		&	& Total	 \\
Yellow   &N($\Delta)$	               & 31365 	&14662		&5873 		&	&51900 \\
\hline
& $n_{\rm ff}^{\triangle}$ 	    &4		&13/3		&14/3		&	& Total	 \\
Red   &N($\Delta)$	    &20270 	&7373		&2135		&	&29778 \\
\hline
& $n_{\rm ff}^{\triangle}$ 	    &5		&16/3		&17/3		&6-7	& Total	 \\
 Gray   &N($\Delta)$	    & 6411 	&3442		&859		&373	& 11085\\
\hline

\end{tabular}
\end{table}

\label{lastpage}
\end{document}